\documentclass[preprint,11pt]{aastex}

\shorttitle{FUSE Translucent Cloud H$_2$ Survey}
\shortauthors{Rachford et al.}

\begin{document}

\title{A FUSE Survey of Interstellar Molecular Hydrogen in Translucent Clouds}

\author{Brian L. Rachford\altaffilmark{1},
Theodore P. Snow\altaffilmark{1},
Jason Tumlinson\altaffilmark{1},
J. Michael Shull\altaffilmark{1,2},
William P. Blair\altaffilmark{3},
Roger Ferlet\altaffilmark{4},
Scott D. Friedman\altaffilmark{3},
Cecile Gry\altaffilmark{5,6},
Edward B. Jenkins\altaffilmark{7},
Donald C. Morton\altaffilmark{8},
Blair D. Savage\altaffilmark{9},
Paule Sonnentrucker\altaffilmark{3},
Alfred Vidal-Madjar\altaffilmark{4},
Daniel E. Welty\altaffilmark{10},
Donald G. York\altaffilmark{10}
}

\altaffiltext{1}{Center for Astrophysics and Space Astronomy, Department of Astrophysical and Planetary Sciences, University of Colorado, Boulder, CO 80309-0389}
\altaffiltext{2}{Also at JILA, University of Colorado and National Institute of Standards and Technology}
\altaffiltext{3}{Department of Physics and Astronomy, Johns Hopkins University, 3400 North Charles Street, Baltimore, MD 21218}
\altaffiltext{4}{Institut d'Astrophysique de Paris, CNRS, 98bis, Boulevard Arago, Paris, F-75014, France}
\altaffiltext{5}{Laboratoire d'Astrophysique de Marseille, B.P. 8, 13376 Marseille Cedex 12, France}
\altaffiltext{6}{ISO Data Center, ESA Astrophysics Division, P.O. Box 50727, 28080 Madrid, Spain}
\altaffiltext{7}{Princeton University Observatory, Peyton Hall, Princeton, NJ 08544}
\altaffiltext{8}{Herzberg Institute of Astrophysics, National Research Council, 5071 W. Saanich Road, Victoria, BC V9E 2E7, Canada}
\altaffiltext{9}{Department of Astronomy, University of Wisconsin, 475 North Charter Street, Madison, WI 53706}
\altaffiltext{10}{Department of Astronomy and Astrophysics, University of Chicago, 5640 South Ellis Avenue, Chicago, IL 60637}

\begin{abstract}
We report the first ensemble results from the {\it FUSE} survey of
molecular hydrogen in lines of sight with $A_V$ $\gtrsim$ 1 mag.
We have developed techniques for fitting computed profiles to the
low-$J$ lines of H$_2$, and thus determining column densities for
$J$ = 0 and $J$ = 1, which contain $\gtrsim$99\% of the total H$_2$.
From these column densities and ancillary data we have derived the
total H$_2$ column densities, hydrogen molecular fractions, and
kinetic temperatures for 23 lines of sight.  This is the first
significant sample of molecular hydrogen column densities of
$\sim$10$^{21}$ cm$^{-2}$, measured through UV absorption bands.
We have also compiled a set of extinction data for these lines of
sight, which sample a wide range of
environments.  We have searched for correlations of our H$_2$-related
quantities with previously published
column densities of other molecules and extinction parameters.
We find strong correlations between H$_2$ and molecules such as CH,
CN, and CO, in general agreement with predictions
of chemical models.  We also find the expected correlations
between hydrogen molecular fraction and various density indicators
such as kinetic temperature, CN abundance, the steepness of the
far-UV extinction rise, and the width of the 2175 \AA\ bump.
Despite the relatively large molecular fractions, we do not see
the values greater than 0.8 expected in translucent
clouds.  With the exception of a few lines of sight, we see little
evidence for the presence of individual translucent clouds in our
sample.  We conclude that most of the lines of sight are actually
composed of two or more diffuse clouds similar to those found
toward targets like $\zeta$ Oph. We suggest a modification in
terminology to distinguish between a
``translucent line of sight'' and a ``translucent cloud.''
\end{abstract}


\section{Introduction}

Molecular hydrogen is the most abundant molecule in the interstellar
medium, dominating the composition of the dense clouds that contain
most of the mass.  Even in diffuse clouds, H$_2$ is present, containing
from 10$^{-6}$ to about half of the total hydrogen nuclei (e.g.,
Spitzer, Cochran, \& Hirshfeld 1974; Shull \& Beckwith 1982).  Clearly,
a full understanding of the physics and chemistry of the ISM requires
a detailed knowledge of molecular hydrogen.

As a homonuclear molecule, H$_2$ has no dipole moment and hence no
allowed rotational or vibrational transitions in the radio and
infrared spectral regions.  With the exception of forbidden quadrupole
transitions in the infrared, which can be seen in emission in regions
heated by radiation or shocks (Timmermann et al. 1996), or, in rare
cases, in absorption
when sufficiently high column densities are probed (Lacy et al.
1994), the only way to observe cold interstellar H$_2$ is through its
electronic transitions in the far ultraviolet.

The Lyman (B$\rightarrow$X) and Werner (C$\rightarrow$X) bands lie in
the spectral region between about 844 \AA\ and 1126 \AA.  The moment of
inertia for this low-mass molecule is small, resulting in
widely separated rotational lines which are easily resolved.  As a result,
the far-UV spectrum of any reddened star is dominated by a wealth of H$_2$
absorption bands.  For a summary of the spectra of these bands, see
Morton \& Dinerstein (1976) or Barnstedt et al. (2000).

Previous far-UV observations of H$_2$ absorption have been conducted by
various short-term missions, including sounding rockets (Carruthers 1970
  -- the first detection of H$_2$ in space); the {\it Hopkins Ultraviolet
Telescope} (e.g., Blair, Long, \& Raymond 1996); {\it ORFEUS} (e.g.,
Barnstedt et al. 2000); and {\it IMAPS} (e.g., Jenkins \& Peimbert 1997).
But by far the most
extensive previous observations of far-UV H$_2$ absorption were
performed by the {\it Copernicus} satellite, which provided the first general
quantitative
studies of interstellar H$_2$ as well as a wealth of information on its
formation, its abundance, and its excitation in space (e.g., Spitzer
et al. 1973; Spitzer, Cochran, \& Hirshfeld 1974; Spitzer \& Zweibel
1974; Jura 1975a,b; for summaries, see Spitzer \& Jenkins 1975 and Shull
\& Beckwith 1982).  The {\it Copernicus} work culminated in a survey of
molecular and atomic hydrogen for 109 stars (Savage et al. 1977;
Bohlin, Savage, \& Drake 1978).

The {\it Far Ultraviolet Spectroscopic Explorer} ({\it FUSE}) instrument is
well suited for observations of the H$_2$ bands in absorption, and studies of
H$_2$ have been a longstanding goal of the {\it FUSE} mission (Moos
et al. 2000).  {\it FUSE}
covers the wavelength region from the Lyman limit at 912 \AA\ to about
1187 \AA\, with spectral resolving power of about $\lambda$/$\Delta$$\lambda$
$\approx$ 20,000.  The throughput of {\it FUSE} has been specifically maximized
in the middle of the Lyman band (around 1050 \AA), providing
by far the most sensitive instrument yet available for observing cold H$_2$
in the ISM.  For a full description of the {\it FUSE} mission and instrumental
performance, see Moos et al. (2000) and Sahnow et al. (2000), respectively.

A subset of the {\it FUSE} Science team is conducting a program
to study H$_2$ in the densest clouds accessible, the so-called ``translucent
clouds'' (van Dishoeck \& Black 1988).  These clouds are defined as being
optically thick (with total extinctions A$_V$ $\gtrsim$ 1 mag) yet still
sufficiently transparent as to allow optical and UV absorption-line
measurements of interstellar atoms, ions, and molecules.  The upper limit
on extinction for a translucent cloud is usually considered to be about
A$_V$ $\approx$ 5 mag; beyond that limit there is usually not enough flux
to allow high-resolution visible-wavelength spectroscopy, much less
ultraviolet spectroscopy, which is always hindered by the rise in dust
extinction at short wavelengths.  Hence the goal of our program has been
to observe and survey molecular hydrogen in translucent clouds, going as
far as possible in the direction of maximum extinction, with the hope of
penetrating clouds with A$_V$ as high as 5 magnitudes.  In doing so, we
planned to extend the {\it Copernicus}-based surveys of Savage et al. (1977)
and Bohlin et al. (1978) to greater extinctions and denser clouds.

More specifically, the goals of the {\it FUSE} translucent cloud H$_2$ 
survey include:

$\bullet$ Measuring total gas column densities, to help in 
determining interstellar gas-phase depletions and chemistry.

$\bullet$ Studying the relationship between H$_2$ and dust extinction,
as an aid in assessing H$_2$ formation models as well as extending
correlations between extinction and H$_2$ column density, sometimes
used to estimate cloud masses.

$\bullet$ Assessing the molecular fraction $f_{\rm H2}$ =
2$N$(H$_2$)/[2$N$(H$_2$) 
+ $N$(H I)] and its relationship to dust extinction and other line-of-sight 
characteristics.

$\bullet$ Measuring gas kinetic temperatures ($T_{01}$) from the 
ratio of $J$ = 1 to $J$ = 0 rotational states, and assessing the corrrelation
of $T_{01}$ with other interstellar quantities.

$\bullet$ Extending direct measurements of the CO/H$_2$ correlation through
UV absorption features of CO.  This correlation
is widely used to assess total masses of molecular clouds, but is 
currently based largely on CO abundances derived from mm-wave
radio observations which do not necessarily sample the same material
as H$_2$ in absorption, and indirect H$_2$ abundances estimated from dust 
extinction measures.

$\bullet$ Assessing cloud physical conditions such as density and 
radiation field intensity by analyzing the excitation of high rotational 
states ($J$ = 2 and greater).

$\bullet$ Comparing H$_2$ absorption measures from {\it FUSE} with infrared 
emission often attributed to complex molecules (e.g., PAHs) as measured by
the {\it IRAS} mission.

$\bullet$ Measuring the abundance of HD in order to infer information 
on the D/H ratio.

As will be seen later in this paper, all but the last three goals have
now been met.  We note also that UV absorption-line
CO data do not yet exist for all of our program stars, so more progress
on the CO/H$_2$ correlation will come later.

Two papers on {\it FUSE} observations of H$_2$ in translucent clouds have 
already been published (Snow et al. 2000 [Paper I]; Rachford et al. 2001
[Paper II]).
In addition to our translucent cloud H$_2$ survey, another subset of the
{\it FUSE} Science Team is conducting a general survey of molecular hydrogen
in diffuse lines of sight, including Galactic stars with relatively
little reddening, stars at high Galactic latitude, Magellanic Cloud
stars (Tumlinson et al. 2002), and H$_2$ in the lines of sight toward
other extragalactic sources such as AGNs (Shull et al. 2000).

The rest of the paper is organized as follows: in \S\ 2 we describe
the criteria for selecting target stars for the FUSE H$_2$ survey, along
with comments on the stars chosen; in \S\ 3 we give details of the
observations and data reduction and analysis; in \S\ 4 we summarize the
results; and in \S\ 5 we discuss these results and their implications.

\section{Target Selection and Stellar Properties}

\subsection{General Criteria for Selecting Target Stars}

Unlike the case of the diffuse H$_2$ survey, which has been able to use
spectra obtained by {\it FUSE} for other programs (e.g., the survey of stellar
winds in OB stars; the survey of hot gas in the Galactic halo; and the
survey of interstellar gas at high latitudes and along lines of sight toward
the Magellanic Clouds and other galaxies), our sole source of data comes
from stars we observe specifically for our program on translucent cloud
lines of sight.  No one observes heavily reddened stars unless
absolutely necessary, because the dust extinction acts to increase the
observing time needed, particularly at short wavelengths.  Thus, we had
to conserve observing time and be
as careful as possible in choosing our target stars while seeking to
fulfill our basic criteria.

The result was a list of 45 stars; 31 assigned to {\it FUSE}
program P116, 4 assigned to Q101, 9 assigned to P216\footnote{As
a historical aside, the P216 program came about once {\it FUSE} had
been in operation for several months, and it was established that
the observing efficiency was better than expected.  PI Team
members were afforded an opportunity to request more guaranteed
time for targets that did not conflict with other programs, and
we selected the additional 9 targets, two of which are included
in the present work.},
and one target (HD 24534) added from the personal program of one
of us (T.P.S.; P193).
Of these 45 stars, 25 had been observed through June 2001, and
23 are included in this paper.

The selection of target stars for the {\it FUSE} translucent cloud H$_2$ 
survey program was based on several criteria:

$\bullet$ Maximizing the total extinction probed;

$\bullet$ Exploring lines of sight known to have a wide range of extinction
characteristics such as $R_V$ (the ratio of total to selective extinction)
and the extinction parameters defined by Fitzpatrick \& Massa (1986, 1988,
1990) on the basis of {\it IUE} data;

$\bullet$ Relative simplicity of line-of-sight velocity structure based
on high resolution observations of \ion{K}{1}, \ion{Na}{1}, and CH;

$\bullet$ Availability of (or feasibility of obtaining) ancillary data
on optical interstellar lines of atomic and molecular species not
observed in the {\it FUSE} passband (including ultra-high resolution
spectra for velocity structure analysis), and the availability (or
feasibility of obtaining) infrared spectra of both gas-phase and
solid-state absorption features.

The first of these criteria was the most critical, due to the trade-off
between maximizing dust extinction and finding stars with sufficient
far-UV flux to be observable with {\it FUSE} within reasonable exposure times.
Thus, we explored known UV fluxes for candidate stars, in most cases having
to extrapolate to shorter wavelengths from the {\it IUE} cutoff at about
1170 \AA .  The extrapolations were based on a combination of the known flux
in the {\it IUE} band, the spectral type and instrinsic flux
distribution of the star, and the shape of the UV extinction curve.

In our initial selection of targets we drew heavily upon the IUE-based
ultraviolet extinction curve survey of Fitzpatrick \& Massa (1986, 1988,
1990) and several papers on optical interstellar absorption lines, both
atomic and molecular (e.g., van Dishoeck \& Black 1989; Gredel, van
Dishoeck, \& Black 1993).  We eliminated stars known to have complex
line-of-sight Doppler
structure due to multiple clouds.   However, we found
that this distinction was almost futile, as nearly every line of sight
turned out to have complex structure when examined at sufficiently 
high spectral resolution.   For example, the star we chose to analyze for
our first paper from this program, HD 73882 (Paper I), turned
out to have no fewer than 21 identifiable Doppler components in the
ultra-high resolution spectra of optical interstellar lines
due to K I and Na I.

One of us (D.E.W.) has led a program to obtain very high-resolution 
optical spectra of our FUSE candidates, and has succeeded in obtaining
spectra at resolving powers of $\lambda$/$\Delta$$\lambda$ $\geq$
150,000 in most cases.  The results will appear in a separate paper
(D. E. Welty et al. 2002, in preparation)
while being made available to us as we analyze
the {\it FUSE} H$_2$ spectra.  But,
these high-resolution spectra are not very important for the current
survey, since we include here only the two lowest rotational states of H$_2$,
whose absorption lines are damped and which therefore can be analyzed
accurately without concern for the detailed velocity structure.  We do,
however, reference the optical spectra in a few cases.

Table 2 provides ancillary data, where available, on carbon-based
molecular species for the stars listed in Table 1.  Additional data
regarding the CH component structure will be discussed in \S\ 5.1.  We
plan to supplement Table 2 by performing our own ground-based observations
of various atomic and molecular species.  We expect especially to emphasize
infrared
spectroscopy in this effort, in order to obtain data on both gas-phase
and solid-state absorption features such as those due to CO, water ice,
and the 3.4-$\micron$ hydrocarbon feature.

\subsection{Extinction properties}
The fundamental quantities used to categorize dust extinction along
a line of sight are $E(B-V)$ and $R_V$ = $A_V$/$E(B-V)$,
where $A_V$ is the total visual extinction in magnitudes.
Since $E(B-V)$ is also expressed in magnitudes, $R_V$ is unitless.
The quantity $E(B-V)$, or color excess, represents the amount of
reddening along a line of sight and gives a measure of the amount
of dust and gas.  The quantity $R_V$, or total-to-selective extinction
ratio, gives a measure of the size of dust grains, and is a convenient
single parameter for describing the overall shape of extinction
curves (Cardelli, Clayton, \& Mathis 1989).  The detailed shape of
the UV extinction curve also carries considerable information.  We
discuss these extinction properties in the following sections.

\subsubsection{$E(B-V)$}
The determination of the $E(B-V)$ color excess is relatively
straightforward for early-type stars.  The intrinsic $(B-V)_0$
colors for O and B stars vary slowly as a function of spectral
type, minimizing errors due to 
inaccurate typing and variations in $R_V$.  Table 3 gives our
adopted values.  In most cases, these values come from either
an analysis of the upper main sequences of OB associations, or
simply from a comparison of the observed colors to tabulated
intrinsic colors for a given spectral type.  The differences between
the various references for a given star are typically only a few
hundredths of a magnitude at worst.  This is comparable to the
differences between different lists of intrinsic colors, and not
much greater than the photometric uncertainties themselves.
In the case of the Be star HD 110432, the value quoted in Table 3
includes a correction to the observed $E(B-V)$ to allow for the
emission line behavior.

\subsubsection{$R_V$}
The situation with regards to $R_V$ is not as simple.  Whatever
the method, measuring $R_V$ requires more difficult observations
with much greater calibration problems.  We will exploit three
methods for estimating $R_V$ in the present work.

If a star does not show excess infrared emission (e.g. Be stars
or other stars with circumstellar material), an extrapolation
of infrared color excesses to infinite wavelength yields an
estimate of $R_V$.  In the method developed by Martin \& Whittet
(1990), the near-infrared extinction curve takes the form,

\begin{equation}
\frac{E_{\lambda-V}}{E_{B-V}} = e\lambda^{-\alpha} - R_V.
\end{equation}

\noindent The infrared extinction in the range 0.9--4.8 $\mu$m has
been expressed as

\begin{equation}
\frac{A_{\lambda}}{E_{B-V}} = e\lambda^{-\alpha},
\end{equation}

\noindent where $e$ is the power-law amplitude, similar to
the $c_4$ parameter of the UV extinction curves described in
\S\ 2.2.3, and $\lambda$ is given in $\mu$m.  Martin \& Whittet
found that the power-law index, $\alpha$, rarely deviates from the
average value of 1.8, even for material with $R_V$ far from the
interstellar average of 3.1.  Thus, to within the normalization
provided by $e$, all extinction curves in the range $\lambda$ =
0.9--4.8 $\mu$m can be described by this universal curve.
With the known quantities $E_{\lambda-V}$/$E_{B-V}$ and 
$\lambda^{1.8}$ as the ordinate and abscissa, respectively,
we can fit a linear relation with slope $e$ and a $y$-intercept of
$-R_V$.  This allows us to apply Eq.\ 1 to cases
where we only have 3 infrared measurements (or even 2
measurements, with much greater uncertainty).  We have verified
that in the cases where we have 4 or 5 measurements, the full
form of Eq.\ 1 gives results consistent with the linearized form;
i.e. $\alpha$ $\approx$ 1.8.  Although Eq. 1 and 2 appear to
hold for photometric bands from $I$ through $M$, we have limited
our fits to the five available bands from $J$ through $M$.  As
Table 3 indicates, we generally had 3 or 4 bands to work with.

One complication of this method is that the colors of the stars
must be corrected for reddening by comparison with
a standard relationship as a function of spectral type.  In
absolute terms, the infrared colors $(J-V)_0$, $(H-V)_0$, etc.
change much more rapidly with changing spectral type than
$(B-V)_0$.  The choice of the appropriate set of intrinsic
colors, as well as the spectral type, leads to significant
differences in the derived value of $R_V$.  Many authors have
used older intrinsic color-spectral type relationships such
as that of Johnson (1966).  However, one must interpolate to
generate the appropriate $H-V$ colors, and the tables are
incomplete for O stars.  More recently, Wegner (1994) compiled
a more complete set of intrinsic colors for O and B stars, and
other similar calibrations exist.

We found infrared photometry in at least 2 filters for 19 of our
23 stars and performed fits to Eq.\ 1 based on intrinsic colors
from both Johnson (1966) and Wegner (1994).
In many cases, photometric uncertainties for a specific star
were not given.  Thus, in our error analysis, we assumed $\sigma$
= 0.03 for $JHK$, $\sigma$ = 0.05 for $LM$, and $\sigma$ =
0.04$E(B-V)$ for $E(B-V)$.  The final statistical uncertainties on
$R_V$ were then close to 0.2 in all cases.

In two cases, HD 110432 and HD 168076, we could not obtain reasonable
fits to Eq.\ 1.  HD 110432 is a Be star, so that we would not expect
this method to work due to the distorted infrared colors.  HD
168076 also shows emission lines.  However, several
other O stars in our sample also show emission lines, yet the
infrared photometry closely follows the expected power law.
We have noticed for this star that the infrared photometry from
Th\'{e} et al. (1990) differs significantly from that given by
Aiello et al. (1988), yet neither set of photometry obeys Eq.\ 1.

For the remaining stars with photometry in 4 or 5 filters and
good fits, the differences between the two sets of intrinsic colors
are large and systematic.  The differences averaged 0.20,
and in all cases the values of $R_V$ were smaller for the Wegner
(1994) intrinsic colors.

For the 7 cases where previous authors have used this or a similar
method (BD $+$31$^{\circ}$ 643, HD 62542, HD 73882, HD 206267, HD 207198,
HD 210121, and HD 210839), our values agree nearly exactly
{\it when we have used the same set of intrinsic colors}.
Considering that the differences in $R_V$ for the two different
sets of intrinsic colors are smallest for the B stars, and that
the Wegner (1994) calibration is much more complete for the O
stars, we have accepted this calibration as the more accurate
overall and give those values in Table 3.

The wavelength of maximum linear polarization generally correlates
well with $R_V$, following the relationship

\begin{equation}
R_V = (5.6 \pm 0.3)\lambda_{\rm max},
\end{equation}

\noindent where $\lambda_{\rm max}$ is given
in $\mu$m (Whittet \& van Breda 1978).  Combined with the uncertainties
in determining $\lambda_{\rm max}$ (a few tenths of a percent to
several percent), this method produces values of $R_V$ accurate to
about 5--10\%, similar to the uncertainties from the first method.

It is worth emphasizing that the relationship
between $R_V$ and $\lambda_{\rm max}$ was calibrated in part from
infrared photometry, so that this second method is not completely
independent of the first.
A further complication is that Eq.\ 3 may not always be satisfied.
Recently, Whittet et al. (2001) found deviations from Eq.\ 3 within
the Taurus dark clouds for a number of lines of sight, in the sense
of larger than normal $\lambda_{\rm max}$ for normal values of $R_V$
derived from IR photometry.

Finally, one can estimate $R_V$ by comparing the ultraviolet
extinction curves to standard curves.  Again, this method requires
calibration from other methods of determining $R_V$.  A comparison
of the shapes of extinction curves can be somewhat subjective.  However,
in an important set of papers, Fitzpatrick \& Massa (1986, 1988, 1990)
devised a six-parameter description of the ultraviolet curves that
could reproduce the entire curve to within the observational errors.
We discuss extinction curves in more detail in the following section,
but the most useful correlation is between $R_V$ and the linear
far-UV rise, the ``$c_2$'' parameter.
The relationship derived by Fitzpatrick (1999) is

\begin{equation}
c_2 = -0.824 + 4.717R_V^{-1}.
\end{equation}

\noindent The typical scatter about this relationship is a few
hundredths in $R_V^{-1}$, or $\sim$0.2 in $R_V$ for small values,
and $\sim$0.5 for large values.  Larger deviations occur for a few
lines of sight.  For example, the extinction curve for HD 62542
has an extremely steep far-UV rise, out of proportion with the
rest of the curve (Cardelli \& Savage 1988).

Unfortunately, in some cases, there are significant disagreements
in $R_V$ derived from the different methods.  However, in most
cases the agreement between the methods is reasonable.  The most
direct method for the determination of $R_V$ is the method involving
IR photometry, and we have such a determination for a majority of our
targets.  We have used this method as our primary source for
$R_V$, using values from polarimetry and the extinction curves in
that order to supplement the IR photometry.  We then simply derive
$A_V$ from the product of $R_V$ and $E(B-V)$ (Table 3).
We have not included specific uncertainties on these values, but
as we noted above, the typical statistical uncertainty on the
photometric values of $R_V$ is 0.2.  The resultant statistical
uncertainties on $A_V$ are then typically 0.1--0.3.  Systematic
effects may be important as noted above in the discussion of the
methods of determining $R_V$.

\subsubsection{Extinction curves}
As mentioned previously, Fitzpatrick \& Massa (1986, 1988, 1990)
devised a six-parameter scheme to describe the UV extinction curve
in the {\it IUE} wavelength range, $\sim$1170--3200 \AA .
The six parameters include three that describe the central wavelength ($x_0$),
width ($\gamma$), and height ($c_3$) of the 2175 \AA\ bump, two that
represent a linear background term ($c_1$ and $c_2$), and one that
describes the far-UV curvature ($c_4$).  With $x \equiv \lambda^{-1}$
in units of $\mu$m$^{-1}$,
the function describing the extinction curves, $k$, is given by

\begin{equation}
k(x-V) = c_1 + c_2 x + c_3 D(x;\gamma,x_0) + c_4 F(x),
\end{equation}

where

\begin{equation}
D(x;\gamma,x_0) = \frac{x^2}{(x^2-x_0^2)^2 + x^2 \gamma^2},
\end{equation}

and

\begin{equation}
F(x) = \left\{ \begin{array}{lr} 0.5392(x-5.9)^2 + 0.0564(x-5.9)^3 & x \geq 5.9\\ 0 & x < 5.9\\ \end{array}\right.
\end{equation}

The function $D(x;\gamma,x_0)$ is the so-called Drude function
(the absorption profile of a damped freely oscillating electron) and
represents the 2175 \AA\ bump.  The far-UV curvature is described
by the function $F(x)$.  As previously noted, $c_2$ correlates well
with $R_V$, and grain size.
Table 4 gives previously published extinction curve parameters
for most of our targets.  In \S\ 4.5 we will investigate various
correlations involving these parameters.

\subsection{Special line-of-sight characteristics}
Many of the lines of sight are of special interest due to their
location and/or environment.  We give a brief overview of each
line of sight in the following sections, including the mention
of particularly interesting values from Tables 1--4.

\subsubsection{BD $+$31$^{\circ}$ 643}
The binary system BD $+$31$^{\circ}$ 643 lies within the cluster IC 348 and
interstellar material associated with the cluster.  The line of sight
is only 8$\arcmin$ from the well studied line of sight toward o
Per.  However, the latter star lies in the background of IC 348
and its associated material (Sancisi et al. 1974).  Snow et al. (1994)
found a ``composite'' extinction curve, with contributions from
material having a broad 2175 \AA\ bump and steep far-UV rise,
and from material having a narrow bump and shallow far-UV rise.
Also, the authors noted very different molecular abundances in
the two lines of sight: enhanced CH and CH$^{+}$ column
densities toward BD $+$31$^{\circ}$ 643, and a reduced CN abundance.
They concluded that star formation in this cluster has altered
the physical and chemical state of the gas.  Snow (1993) has
argued that the enhanced CH$^{+}$ column density is larger
than would be expected from shock formation, and proposed a
radiative formation mechanism due to a very strong UV radiation
field.

BD $+$31$^{\circ}$ 643 is surrounded by a circumstellar disk, only the
second such disk ever directly imaged (Kalas \& Jewitt 1997).
Andersson \& Wannier (2000) suggest that more then one magnitude
(approximately half) of the total visual extinction may arise in
this disk.  This might lead to a composite extinction curve, and
a single $R_V$ might not be an appropriate description of the
line of sight.

\subsubsection{HD 24534}
This Be star is more commonly known as X Persei and has an
X-ray-bright pulsar companion.  However, the line of sight toward
the pair intersects a molecular cloud, and the UV flux is
among the largest known for a target with $A_V$ $>$ 1 mag.
In fact, it was (barely) bright enough to be observed with
{\it Copernicus} (Mason et al. 1976).  They found log $N$(H$_2$)
= 21.04, the largest such value obtained with {\it Copernicus}.
The molecular fraction was also estimated to be very large (Snow
et al. 1998), albeit the H$_2$ column density was rather uncertain.

\subsubsection{HD 27778}
This binary star lies within or just behind the Taurus-Auriga molecular
cloud complex (Kenyon, Dobrzycka, \& Hartmann 1994), sampling
relatively low-extinction material in this complex (Whittet et al.
2001).  Although fit parameters have not been published
for the extinction curve, the curve depicted in Joseph
et al. (1986) indicates greater-than-normal far-UV extinction,
and a shallower-than-normal 2175 \AA\ bump, characteristic of
a ``dense cloud'' environment.

We note that while Joseph et al. (1986) reported an \ion{H}{1}
column density log $N$ = 21.59$\pm$0.11 from measurement of the
Ly$\alpha$ line, at spectral type B3 V the Lyman-series
interstellar lines will be strongly contaminated by stellar lines
(Savage \& Panek 1974; Shull \& van Steenberg 1985; Diplas \& Savage
1994).

\subsubsection{HD 62542}
HD 62542 lies in or just beyond the Gum nebula, and the line
of sight passes through an isolated ridge of molecular material
(Cardelli et al. 1990; O'Donnell, Cardelli, \& Churchwell 1992).
Despite a normal value of $R_V$, Cardelli \& Savage (1988) found
that the UV extinction curve is peculiar in two important respects.  
The far-UV portion of the curve is unusually steep, and the
2175 \AA\ bump is weak and strongly shifted to 2110 \AA .
Combined with the following additional observations, Cardelli
et al. (1990) concluded that the line of sight contains very high
density material.
First, optical maps of the Gum nebula complex indicate small,
dark globules scattered through the ridge of material through
which the light from HD 62542 passes.
Second, the CN excitation temperature is considerably larger than
the temperature of the cosmic background radiation, indicating
significant collisional excitation.
Finally, the column densities for CN and CH are exceptionally
large given the relatively small visual extinction.
The conclusion reached by Cardelli et al. (1990) was a remarkably
high density of $n$ $\sim$ 10$^{4}$ cm$^{-3}$, and a cloud with
thickness $d$ $\sim$ 0.2 pc and mass $m$ $\sim$ 1 M$_{\sun}$.

Other density diagnostics suggest less extreme conditions, i.e.,
C$_2$ excitation implies $n$ $\sim$ 500--1000 cm$^{-3}$ (Gredel
et al. 1991, 1993), and CN observations also imply $n$ $\sim$
500 cm$^{-3}$ (Black \& van Dishoeck 1991).  With this value,
the cloud size is larger,
i.e., 0.5--1.0 pc (Snow et al. 2002).  In any case, the presence
of such high densities in a cloud with relatively small extinction
is unusual.

Despite the large column densities of CN and CH, the CH$^{+}$ column
density is remarkably small.  If shocks are the main formation
mechanism for CH$^{+}$, this result would be consistent with
high-density material.  However, as discussed in \S\ 4.1 the
formation mechanism of CH$^{+}$ is still in question.

\subsubsection{HD 73882}
Molecular hydrogen has been studied in some detail for this
line of sight in Paper I.  We take advantage of
additional data obtained subsequent to this paper, but this
information does not alter the original conclusions.
The extinction is among the largest in the present study.
The extinction curve shows a shallow 2175 \AA\ bump with
steeper than normal far-UV extinction, characteristics common
to ``dense cloud'' curves.  However, the overall conclusion of
Paper I was that the line of sight appeared to be
similar to diffuse clouds, and may contain a collection of
several clouds.

\subsubsection{HD 96675, HD 102065, HD 108927}
These three targets all lie within or just beyond the Chameleon
dark clouds.  Although the color excesses are among the smallest
of the present sample, the column density of the CH radical
toward HD 96675 and HD 102065 suggests a considerable molecular
content (Gry et al. 1998).  High-velocity atomic gas is also
seen toward these two stars (Gry et al. 1998).  Gry et al. (2002)
have studied H$_2$ toward these lines of sight and concluded
that a component of the high-$J$ excitation comes from collisional
excitation in regions of warm gas.

Despite their relative proximity, the three lines of sight sample
material with strongly differing dust properties, as indicated
by {\it IRAS} photometry, the UV extinction curves, and $R_V$
(Boulanger et al. 1994).
Unfortunately, the methods for the determination of $R_V$ given
in Table 3 show exceptional disagreement.

\subsubsection{HD 110432}
This line of sight has been studied in detail in Paper II.
The main conclusion was that the line of sight was similar to
$\zeta$ Oph, based on the column densities in the $J$ = 0--7
levels, as well as chemical modeling.
The CN column density is quite small relative to our present sample,
but is comparable to that seen toward $\zeta$ Oph.

\subsubsection{HD 154368}
This star is one of the few with sufficient flux to be observed
by {\it FUSE} through 2.5 magnitudes of visual extinction.  The
component structure appears to be dominated two components, and many
species have been well studied with {\it HST} (Snow et al. 1996).
The extinction curve is more or less normal, but tends
slightly towards the ``dense cloud'' curves (Snow et al. 1996).
They concluded that the line of sight contains extended regions
of moderately dense gas, as opposed to one or more dense cloud
cores.

\subsubsection{HD 167971}
This target is actually an eclipsing binary pair, with a third
supergiant component which is the most luminous member of the group
(Leitherer et al. 1987).  The three stars may represent a Trapezium-type
system or possibly a chance alignment.
All three stars are likely members of the Serpens OB2 association,
and at least the supergiant star lies within the cluster NGC 6604.
Serpens OB2 is physically related to the \ion{H}{2} region Sharpless 54.
This target has the largest color excess and visual extinction of
the present sample.

\subsubsection{HD 168076}
This star lies within the cluster NGC 6611, which in turn
is associated with the dusty \ion{H}{2} region M16.  At type O4,
this star is the earliest target included in our survey.
The cluster stars in NGC 6611 show significant intracluster and/or
foreground differential reddening (Orsatti, Vega, \& Marraco 2000).
These authors found evidence for a slightly larger value of
the wavelength of maximum polarization, $\lambda_{max}$, than in
the general ISM.  However, the line of sight toward HD 168076 did
not conclusively show this effect.

\subsubsection{HD 170740}
Although a bright target in both the optical and ultraviolet, there
have been relatively few studies of this line of sight.  The column
densities reported in Table 2 are unremarkable and the UV extinction
curve was judged normal by Witt, Bohlin, \& Stecher (1984).

\subsubsection{HD 185418}
This line of sight is not well studied.  Fitzpatrick \& Massa
(1986, 1988, 1990) included the target in their extinction curve survey.
The extinction curve shows a stronger than normal 2175 \AA\ bump,
along with less than normal far-UV extinction.

\subsubsection{HD 192639}
HD 192639 lies within the cluster NGC 6913.
Nichols-Bohlin \& Fesen (1993) studied the complex interstellar
environment centered on the Wolf-Rayet star HD 192163, lying
just one degree away from HD 192639 and at nearly the same
distance.  The authors found evidence for a supernova shell
surrounding HD 192163, and perhaps HD 192639.  Most of the lines
of sight toward hot stars in the vicinity showed high-velocity
components (primarily at negative velocities) including HD 192639,
indicative of the multiple superbubble structure surrounding the
Cyg OB1 and OB3 associations.

P. Sonnentrucker et al. (2002; in preparation) have undertaken
a detailed study of abundances along this line of sight, combining
the {\it FUSE} data with high-resolution {\it HST} spectroscopy.
Based on the inferred low particle density, complex component
structure, and chemical depletions, they concluded that the line
of sight is a collection of diffuse clouds.

\subsubsection{HD 197512}
This is another line of sight from the Fitzpatrick and Massa
(1986, 1988, 1990) sample that otherwise has not been well studied.
The extinction curve shows greater extinction than the standard
curve at all wavelengths, suggesting a possible miscalibration
with $E(B-V)$ or spectral type of the comparison target.

\subsubsection{HD 199579}
This is the second target in the present survey for which
$N$(H$_2$) was also derived from {\it Copernicus} observations.
The star contributes to the excitation of the North America
Nebula (NGC 7000), but is not the primary excitation source
(Neckel, Harris, \& Eiroa 1980).  The extinction curve shows
larger than normal far-UV extinction, but the 2175 \AA\ bump
is normal.  The line-of-sight CN column density is unusually
small.

\subsubsection{HD 203938}
The extinction curve for this line of sight is similar to
that for HD 197512 in the sense that the extinction is
greater than normal at all wavelengths (Fitzpatrick \& Massa
1990).  However, the effect is not as great for this target.
Otherwise, the line of sight is not well studied.

\subsubsection{HD 206267}
HD 206267 is a Trapezium-like quadruple star system (Abt 1986) within
the cluster Trumpler 37.  The cluster is associated with the \ion{H}{2}
region IC 1396, and the hottest star of the group, HD 206267A, is the
main exciting source for the region.  The {\it FUSE} observation contains
contributions from components A and B.  The cluster and \ion{H}{2}
region have been well studied.  Morbidelli et al. (1997) found
relatively uniform extinction across the cluster both in terms of
total extinction and the total-to-selective extinction ratio.
Clayton \& Fitzpatrick (1987) derived UV extinction curves and found
that HD 206267AB, like most early-type members of the cluster, shows
a normal 2175 \AA\ bump and stronger than normal far-UV
extinction.  The uniformity of extinction and extinction curves
for the cluster stars led to the interpretation that most of the
extinction toward HD 206267AB is foreground with a small contribution
from IC 1396 (Clayton \& Fitzpatrick 1987; Morbidelli et al. 1997).

More recently, however, variations in the column densities of individual
velocity components over spatial scales of $\sim$10,000--40,000 AU
have been observed toward the HD 206267 system (Lauroesch \& Meyer 1999;
Pan, Federman, Welty 2001).  The most striking variations appeared
for CN, the most density sensitive of the species studied.

\subsubsection{HD 207198}
The extinction curve derived by Jenniskens \& Greenberg (1993)
indicates much stronger than normal far-UV extinction, and the
other indicators of $R_V$ also indicate that this is an unusual
line of sight.  In this regard, it is similar to the line of
sight toward HD 210121, but less extreme (see below).

\subsubsection{HD 207538}
The polarization data suggest that this line of sight may be
similar to HD 207198 and HD 210121, with a small value of $R_V$.
However, the line of sight is otherwise poorly studied,
and we are unaware of a published extinction curve that might
verify the unusual extinction.

\subsubsection{HD 210121}
HD 210121 lies within or behind the remarkable high-latitude molecular
cloud DBB 80 (D\'{e}sert, Bazell, \& Boulanger 1988; de Vries \& van
Dishoeck 1988), and this is the only
high-latitude line of sight in the present survey.  The UV extinction
curve exhibits one of the steepest far-UV rises ever observed
(Welty \& Fowler 1992), consistent with the exceptionally small
total-to-selective extinction ratio, $R_V$ = 2.1.  In addition,
the 2175 \AA\ bump is very weak, such that all available extinction
parameters suggest dense material.  This line of sight is
especially important as it is the only one known with such extreme
extinction that is bright enough for {\it FUSE} observations.  While
HD 62542 has a similar far-UV rise, it does not have the corresponding
extreme $R_V$.
Welty \& Fowler (1992) also found a small ratio between the 100 $\mu$m
flux and the total hydrogen column density, suggesting a smaller than
average incident radiation field consistent with the cloud's
location 150 pc from the Galactic plane.

Grain models suggest an excess of small grains and a deficiency of
large grains relative to the average interstellar medium (Larson et
al. 2000).  Li \& Greenburg (1998) have modeled the extinction,
polarization, and emission in the molecular cloud by including
a contribution from grains which have undergone erosion and thus
have thinner mantles.

\subsubsection{HD 210839}
Also known as $\lambda$ Cephei, this is the third and final target
in our list for which $N$(H$_2$) was also derived from {\it Copernicus}
observations.  As the optically brightest target in the present
survey, many ground-based studies have been performed.  However,
the column densities of the various species (Table 2) are
generally normal.  Bless \& Savage (1972) published a partial UV
extinction curve out to 1800 \AA\ which indicates a nearly normal
2175 \AA\ bump.

Jenkins \& Tripp (2001) found two main components in high resolution
observations of \ion{C}{1} lines (with possible structure within each
component).  The two components were separated by about 23 km s$^{-1}$.
The weaker red component, presumably associated
with the expanding Cepheus bubble, contains high-pressure gas
($p$/$k$ = 10$^{4.8}$ cm$^{-3}$ K), as derived from \ion{C}{1}
and \ion{O}{1} fine-structure excitation.  This component would
be a good candidate for the presence of H$_2$.
While the 23 km s$^{-1}$ separation could be resolved in weak H$_2$
lines in our {\it FUSE} spectra, saturation broadening in lines with
$J$ $\leq$ 4 (and the strongest lines with $J$ = 5) overwhelms any
potential weak red-shifted component.
In the weaker lines with $J$ = 5--7 we do not see resolved component
structure.  Based on the strong CH (and CH$^{+}$) absorption seen
by Crane, Lambert, \& Sheffer (1995) at the same velocity as the blue
components of
\ion{C}{1} and \ion{O}{1} seen by Jenkins \& Tripp (2001), we assume
that all the observed H$_2$ is also associated with this component.

\section{Observations and data analysis}
Table 5 gives information on our {\it FUSE} observations.  This
survey is based on a total of 288 ksec of integration time on
23 targets, taken over the course of 20 months.  Our targets are
generally relatively faint, and only HD 24534, HD 110432, HD 170740,
HD 199579, and HD 210839 were bright enough to require spectral-image
mode (Moos et al. 2000).

BD+31$^{\circ}$ 643 and HD 73882 were observed during
the ``early release observation'' phase of the mission, and were
subsequently reobserved due to incomplete data coverage.  The first
half of the initial observation of HD 207538 was lost due to a software
problem onboard the spacecraft, and this target was also reobserved.
In several cases, the steepness of the extinction curve and/or
the faintness of the target prevents us from having usable
data in the shortest wavelength segments (SiC 1B and 2A).  However,
the wavelengths below 980 \AA\ are not used in the present H$_2$
analysis.

As we are fitting very broad profiles and our earliest datasets
have poor S/N, we have not made an attempt to re-process all of
the data sets with the most recent version of the CALFUSE pipeline.
We have, however, applied the most appropriate version of the wavelength
calibration to each dataset.

In all cases, each observation is broken into two or more individual
integrations.  Before combining the individual spectra, we perform
a cross-correlation analysis on a cluster of narrow lines near
the center of each data segment to co-align the spectra.  Since
the spectra are highly oversampled (one resolution element
corresponds to about 9 pixels), we only applied integral pixel
shifts, so that the data are never interpolated or resampled.
The pixel-by-pixel uncertainties are propagated through the
co-additions, and each of the 8 detector segments is processed
separately.  Significant differences in flux calibration, line
spread function, and wavelength solutions conspire to make useful
co-additions of data from different segments difficult.
Despite these differences, we do find generally
good agreement between segments; however, the differences are
larger than the formal uncertainties within each segment, as
discussed in the following section.

We have reported per-pixel values of S/N in Table 5.  With the
typical 9 pixel resolution element, the S/N for one resolution
element could be 3 times as large.  The true S/N for a resolution
element for our highest quality observations may be limited by
fixed-pattern noise and the lack of a flat-fielding correction.
However, thermal motions present in the mirrors and gratings help
to dither the spectral image across the detectors over the course
of multiple integrations and help smooth out the fixed-pattern
noise.  Thus, in most cases our estimated S/N for a resolution
element is close to the optimum value.
More information concerning the on-orbit performance of {\it FUSE}
is given by Sahnow et al. (2000).

\subsection{Data Analysis}
\subsubsection{Overview of the problem}
A description of our analysis procedures is given in Paper II.
However, in the present paper we expand upon several
issues involved with fitting these complex spectra.  This discussion
is based on a long period of experimentation with various analysis
techniques on spectra with a wide range of S/N and column densities.
These techniques were also used by Tumlinson et al. (2002), with
appropriate modifications to allow for the presence of both Galactic
and Magellanic Cloud components in the overall H$_2$ spectrum.

Figure 1 shows sample spectra for a high S/N target (HD 210839)
and a low S/N target (HD 154368).  The equivalent widths of the
undamped $J$ $\geq$ 3 lines can generally be measured individually
without regard to the specifics of the unresolved component
structure along the line of sight, and then a curve of
growth analysis can be performed.  We are performing such an
analysis, and these results will be presented at a
later date.  However, to determine the total H$_2$ column density
we must perform profile fits on the highly damped $J$ = 0 and 1
profiles.  In addition, since the R(2) lines are blended with the
main $J$ = 0 and 1 profiles, we must include $J$ = 2 in the profile
fits.  Fortunately, the P(2) lines are sufficiently isolated to
constrain the $J$ = 2 column densities for our present purposes.

The $J$ = 0 and 1 lines themselves are too heavily damped to be
sensitive to the detailed component structure (or $b$-value if a single
absorbing cloud is assumed).  The $J$ = 2 lines are somewhat
sensitive to this choice, and in turn the blending between the P(1) and
R(2) lines can affect the derived $J$ = 1 column density.
However, we have found that a change in $b$-value of a factor
of 2 typically changes log $N$(1) by just 0.01 dex, even with a much
greater change in $N$(2).

The steep extinction curves in these heavily reddened lines of
sight and the overlap between Lyman and Werner bands of
H$_2$ and the Lyman \ion{H}{1} lines prevent us from reliably
using H$_2$ bands near and shortward of Ly$\beta$.  Thus, we
have considered Lyman bands of H$_2$ from (0,0) through (4,0).

The continuum radiation from the background hot stars
is punctuated by photospheric lines.  If these lines lie in the far
wings of the damped profiles, they can be easily divided out of the
spectra.  If they lie near the zero-intensity cores of the profiles,
these lines do not affect the overall H$_2$ spectrum.  However, 
contamination of portions of the H$_2$ profiles with normalized
intensities of $\sim$0.5
is both difficult to remove and can cause significant disruption
of the H$_2$ spectrum.  In Paper II we noted that fits of the
Lyman (0,0) band in HD 110432 exhibited such contamination, and that
the (3,0) band fits were also possibly affected.  Further exploration
of this issue in additional spectra indicate that contamination of
these bands is common.  While detailed modeling of the stellar
spectra is beyond the scope of this work, we note that the (1,0),
(2,0), and (4,0) bands appear to be the cleanest of the long
wavelength bands in hot star spectra observed through less H$_2$.
In the interest of producing a uniform measurement of H$_2$
column densities across our sample, we are limiting our $J$ = 0
and 1 analysis to these bands.  These bands appear a total of
nine times on five different detector segments.

A final problem is the wide range in data quality.  As seen in Table
5, the per-pixel S/N varies from unity to nearly 30.  Our data analysis
techniques have to work well across this large range.  

\subsubsection{Fitting techniques}
Our goal is to match a model spectrum of the low-$J$ lines to the
data by varying the $J$ = 0, 1, and 2 column densities, a quadratic
continuum, and a zero-point wavelength shift.
We have applied two distinct fitting methods to our data to minimize
the squared difference between the model and the data, non-linear
least squares (i.e., the Levenburg-Marquardt ``CURFIT'' algorithm from
Bevington \& Robinson 1992), and the ``downhill simplex method'' (the
``AMOEBA'' algorithm from Press et al. 2000).  The non-linear least
squares method has the advantage of producing formal uncertainties
on each fit parameter from the covariance matrix when using the
appropriate weighting scheme.  In addition, it is much less
computationally intensive.  However, this method also requires the
evaluation of partial derivatives with respect to each parameter,
but our modeled profiles do not have analytical derivatives.  The
downhill simplex method works to minimize a quantity, in this case the
difference between the model and the data, weighted in some manner.
The method is more computationally intensive but only requires
function evaluations and not derivatives.  It can also be much more
robust, particularly when dealing with a large number of fit
parameters.

We have experimented with several weighting schemes for the data,
including ``instrumental'' (1/error$^{2}$), ``signal-to-noise''
(data/error), and ``uniform''.  Instrumental weighting has the
advantage of producing genuine $\chi^2$ values and appropriate
statistical error bars on each fit parameter, and is the usual
choice for astronomical data.  However, we have many data
sets with very poor data quality (per-pixel S/N of a few or
less).  In these cases, the generally Poissonian errors can not be
reasonably approximated by a normal distribution.  This fact tends
to skew the weighted fits toward the pixels with smaller values
instead of bisecting
the data points.  Thus, in the poorest quality observations in the
present sample, the instrumentally weighted fits do not provide
a good match to the data.  If we re-bin the data to increase the S/N,
this becomes less of a problem.  However, such re-binning has its own
set of problems.  Signal-to-noise weighting is somewhat less
susceptible to this problem, although this weighting method is rarely
used.  A uniformly weighted fit removes this effect completely, but
gives limited information on the uncertainties of the fit parameters.
In the uniformly weighted fit we are in effect assuming that the
perfect model should bisect the data points in
a given range for any S/N.

We have tested the various combinations of fitting techniques and
weighting schemes.  As an example, in Table 6 we show the results
of a suite of fits of the (1,0), (2,0),
and (4,0) bands from the LiF 1A and LiF 2A segments for HD 199579.
This is a high-quality spectrum with S/N of 20 per-pixel at the
maxima between the H$_2$ bandheads, and we would expect to see good
agreement among the various methods.  The values in Table 6 show
that this expectation is verified.  Thus, we can choose the fitting
technique and weighting scheme most appropriate for our poor-quality
data without sacrificing the accuracy of the fits to the good-quality
data.

As we observed in Paper II, the largest variation in the fit
parameters occurs when comparing one band to another.  In addition,
the formal uncertainties we derive for the high-quality data sets
through both the CURFIT method are less than 0.01 dex.  We also
performed Monte Carlo simulations whereby we either added noise to
synthetic profiles matching the data, or added additional noise to
the data itself, and again the changes in the column densities were
much smaller than the band-to-band and segment-to-segment differences.
From these findings, we confirm our conclusion in Paper II that the
largest source of errors in the low-$J$ column densities
are from effects other than Poisson noise.  These effects include
low-level contamination from stellar lines, fixed-pattern noise,
wavelength-dependent errors in the flux calibration, and other factors.

\subsubsection{Summary of the fitting technique}
Based on the previous discussion we now describe our revised
fitting technique in full.
Before fitting the spectra we identify all visible atomic lines
and all visible H$_2$ lines with $J$ $\geq$ 3, model them with Gaussian
profiles, divide the lines out of the spectrum, and exclude from
further fitting the cores of the removed lines if they dropped
below half of the local continuum level.  We also remove obvious stellar
lines with a similar procedure.  This leaves only seven fit parameters:
a quadratic polynomial continuum, logarithmic column densities for
$J$ = 0, 1 and 2, and a zero-point wavelength shift.  In Paper II,
we included the high-$J$ and atomic lines in our fits instead of an
outright removal of the lines.  Our choice in this matter has
little effect on the quality of the fits, but the reduction in fit
parameters if we remove the lines reduces the computational time,
and tends to make CURFIT as robust as AMOEBA.  Also, the number of
individual lines that must be modeled is reduced, again reducing
the computational time.
During the line-removal phase, we also select the appropriate
wavelength ranges for the fitting of each band.  We make these
ranges as uniform as possible from target-to-target and for fits
of the same band in different detector segments of the same
observation.

We assign a single $b$-value to represent the overall component
structure, which only affects the modeling of the $J$ = 2 lines.
As stated previously, this value need not be very accurate, and we
have used preliminary values from our curve-of-growth analysis of
the high-$J$ lines, combined with the high-resolution ground-based
data described in \S\ 1.

While there must be unresolved velocity structure in nearly all
cases, we do not see evidence for resolved structure in
H$_2$ in any of the 23 targets.  The velocity separation we
could detect varies from target to target due to S/N issues
and differences in strength and saturation level of the
high-$J$ lines.  However, typical values are 20--30 km s$^{-1}$.

We model the line-spread function of the spectrograph with a
Gaussian of FWHM corresponding to a resolution, $R$ $\approx$
17,000.  This corresponds to the typical resolution of our spectra
which were all observed through the largest available slit
(30$\arcsec$ $\times$ 30$\arcsec$).  In many cases we are achieving greater
resolving power, but even the $J$ = 2 lines are considerably
broadened beyond the instrumental profile, and the $J$ = 0 and 1
profiles are not affected by the choice of the line-spread
function for any reasonable value.

The H$_2$ model itself includes the R(0), R(1), P(1), R(2), and
P(2) lines of the band being fitted, as well as the R(0), R(1), and
P(1) lines of adjacent bands.  We must include the latter lines to
account for the overlapping damping wings of the $J$ = 0
and 1 lines from adjacent bands.  We use line parameters from
Abgrall et al. (1993).  Once this
large model spectrum is calculated on a somewhat finer wavelength
grid than the actual spectrum, it is convolved with the line-spread
function, the zero-point shift is applied, and the model spectrum
is rebinned to the grid of wavelengths from the actual spectrum.

The final fits we report use the CURFIT routine with uniform weighting.
In some cases the data quality is too poor at the shorter wavelengths
and we can not adequately perform fits of the (4,0) band.  Also, the
SiC channels have poorer S/N in our range of interest, so in some
cases we can obtain fits in the LiF channels but not SiC.  A more
subtle problem also occurs with the (4,0) band in certain cases.
Considerable information is carried by the ``bump'' between the cores
of the R(1) and P(1) lines.  This small non-zero section in the
spectrum at the saturated core of the vibrational bandheads is very
sensitive to the $J$ = 1 column density, and somewhat sensitive to the
$N$(1)/$N$(0) ratio, yet it is usually weak enough to be insensitive to issues
such as continuum placement.  Due to the smaller $f$-values of the
lines in the (2,0) and (1,0) bands, the bump is very prominent.  But
in the (4,0) band, a combination of large column density, poor data
quality, and the larger $f$-values can totally eliminate the bump.
This situation leaves the fitting routine with only the blue and red wings
of the overall $J$ = 0 and 1 profile to constrain the column densities
and still accurately define the continuum.  In some cases, this leads
to unreasonable fits where the continuum will be strongly
parabolic instead of relatively flat, and the column density may
disagree with other bands by factors of 5--10.

\subsection{Error analysis}
In the ideal case we have 9 independent measurements of the
column densities per observation.  (For HD 73882 and HD 206267,
the multiple observations give us more measurements.)  Since
the formal errors on the fits are smaller than the fit-to-fit
differences, we adopt the same error analysis as in Paper II,
and use as 1-$\sigma$ errors the sample standard deviation
of the individual fits.
For this choice to be appropriate, the individual fit parameters
must be more-or-less normally distributed.  We have 16 datasets
for which we could obtain all 9 possible column density
measurements, and we have used these datasets to search for
systematic differences between the individual band/segment
combinations.  We determined the quantity

\begin{displaymath}
d = \frac{\log N(J)_i - \langle\log N(J)\rangle}{\sigma_{n-1}}
\end{displaymath}

\noindent for each of the measurements; i.e., the normalized deviation 
from the line-of-sight mean.  If there are no systematic differences,
the distribution of $d$ should be normal.  Then we found the
average of the 16 $d$-values for each band/segment
combination, along with the error of the mean.

Table 7 gives the results of this analysis, along with
the results for the same band in all segments, and all
bands in all segments (whose average must be zero).
We immediately see that there are a few systematic
differences.  The largest difference occurs for the LiF 1A
(2,0) $J$ = 0 fits which average nearly 1.5 standard deviations
below the overall mean.  The red wing of this profile is
at the very edge of the detector segment which may affect
the continuum determination.  There are also relatively large
systematic effects in the positive sense for both (1,0) band
fits for $J$ = 0, which effectively cancels out LiF 1A (2,0)
in the overall average.  The source of these effects is
consistent with the presence of a very weak stellar line in
the blue wing of some of the (1,0) profiles, and an inspection
of the fits does suggest that this is the case.  In contrast
with $J$ = 0, the $J$ = 1 values show much more
subtle differences.

Overall, it appears that while systematic differences do indeed
occur, they do not have a large effect on the overall
results.  For example, if we were to exclude the LiF 1A (2,0)
fits from the averages, the logarithmic column densities would
only change by a few hundredths.  We also note that since
we have excluded the (3,0) band we do not see band-to-band
differences as large as those reported in Paper II.  Interestingly,
while the cases where we have the fewest individual measurements
involve bands with large systematic differences (i.e., LiF 1A (2,0),
LiF 1A (4,0), and LiF 2A (1,0)), these systematic effects almost
exactly cancel each other out when looking at the ensemble for
$J$ = 0, and are small in any case for $J$ = 1.  For these reasons,
we have included all of these bands in the final averages and
uncertainties, i.e. the mean and sample standard deviation
of the column densities from the (up to) 9 fits.

\section{Results}
Table 8 summarizes the measured and derived quantities relating
to our H$_2$ observations, which we will discuss individually in
the following sections.  We have generally used atomic hydrogen
column densities from the literature, derived from {\it IUE}
observations of the Ly$\alpha$ line.  In two cases, we report
a new determination of $N$(\ion{H}{1}) from our own profile fits
of {\it IUE} data.  For stars later than about spectral type
B2, the contribution from the stellar Ly$\alpha$ line becomes
large enough to seriously contaminate the interstellar line
(\S\ 2.3.3).
Thus, for six lines of sight we have estimated $N$(\ion{H}{1})
from the relationship between $E(B-V)$ and $N$(H$_{\rm tot}$) (see
\S\ 4.2).

Our fundamental observed quantities are $N$($J$=0) and $N$($J$=1),
from which we can derive the total molecular hydrogen column
density (\S\ 4.1), total hydrogen column density (\S\ 4.2), kinetic
temperature (\S\ 4.3), and hydrogen molecular fraction (\S\ 4.4).
In addition, we can assess correlations between H$_2$ parameters
and extinction curve parameters (\S\ 4.5).
The plots in these sections include lines of sight for which
$N$(H$_2$) was measured by {\it Copernicus} (Savage et al. 1977).
For these targets, we used $N$(\ion{H}{1}) from Diplas \& Savage
1994 ({\it IUE}) and Bohlin et al. 1978 ({\it Copernicus}),
in order of preference, with most values coming from the former.

\subsection{Molecular hydrogen column density}
While most of our targets have never been observed at moderate resolution
in the far-UV, three targets in our present program were observed by
{\it Copernicus}, providing measurements of $N$(0) and $N$(1),
or at least $N$(H$_2$).  Table 9 compares those values with our
new values.  The difference for $N$(1) for HD 24534
is quite large, but on the whole the differences are reasonable,
given the much larger uncertainties on the {\it Copernicus}
measurements due to poorer S/N.  We also note that we have analyzed
several {\it Copernicus} targets with our fitting techniques and find
very close agreement with the published values.

With the exception of the uncertain measurement of $N$(H$_2$)
toward HD 24534 (X Per) by Mason et al. (1976), which we have refined,
all but six of our present H$_2$ column densities are larger than any
observed with {\it Copernicus}.  In four cases, our column density is
larger than the revised value for X Per.  We have thus provided the
first significant sample of lines of sight with log $N$(H$_2$) $\approx$ 21.

Figure 2 shows $N$(H$_2$) as a function of color excess for
log $N$(H$_2$) $>$ 20.  Although the column density appears to
level off at large color excess, much of this leveling is due to
our semi-logarithmic presentation and the linear relationship
between total hydrogen column density and color excess described
in the following section.
Interestingly, most of the scatter in $N$(H$_2$) for large color
excess is due to scatter in $N$(0), as $N$(1) remains nearly constant
at 10$^{20.5}$ cm$^{-2}$.  For the 14 {\it FUSE} targets with
$E(B-V)$ $>$ 0.4, the standard deviation of $N$(0) is 0.21 dex as
compared with 0.11 dex for $N$(1), and 0.15 dex for $N$(H$_2$).  This
finding is unchanged if we only consider values with uncertainty of
0.1 dex or less.  However, it is not clear if this apparent
threshold at 10$^{20.5}$ cm$^{-2}$ is meaningful.

Given our new H$_2$ measurements, we can extend the range of
column densities in exploring correlations with other molecules.
In Figure 3 we show the relationships
between $N$(H$_2$) and $N$(CH), $N$(CH$^{+}$), $N$(CN), and $N$(CO).  For
the {\it Copernicus} H$_2$ targets we took column densities from the
compilations of Welty \& Hobbs (2001) and Federman et al. (1994)
for CH, Allen (1994) for CH$^{+}$, and Federman et al. (1994) again
for CN and CO.  We have only included {\it absorption}
measurements since that step gives the best guarantee that we are
sampling the same material as in the H$_2$ measurements.

The excellent relationship between CH and H$_2$ seen in previous
investigations continues to be reflected when adding our new data.
The relationship is nearly linear, in agreement with Danks, Federman,
\& Lambert (1984).  Chemical models predict a linear relationship between
$N$(CH) and $N$(H$_2$),
with some scatter due to variations in density (Danks et al. 1984;
van Dishoeck \& Black 1988, 1989).

We also find a linear relationship between CH$^{+}$ and H$_2$,
with some outlying points and considerable scatter.  The primary
formation reaction for CH$^{+}$ is endothermic, and thus shocks
have been proposed as an energy source for the reaction (Elitzur
\& Watson 1978).  Lambert \& Danks (1986) found a good correlation
between ``warm'' gas as measured by the rotational excitation of
H$_2$ for $J$=3--5, and $N$(CH$^{+}$), supporting the shock
hypothesis.  Our future measurements of
rotationally excited H$_2$ may shed additional light on this issue.

Allen (1994) previously found that log $N$(CH$^{+}$) increases with
$E(B-V)$ up to about 0.6, then levels off.  However, this primarily
occurs due to the semi-logarithmic axes and this relationship
actually remains more or less linear up through $E(B-V)$ = 1.2.  Gredel
(1997) also found linear relationships between $N$(CH$^{+}$) and extinction
($A_V$) within individual OB associations.  In addition, Gredel
found a correlation between $N$(CH$^{+}$) and $N$(CH).  Gredel concluded
that the dissipation of turbulence may be an important production
mechanism.

Another statistically significant relationship appears between
CN and H$_2$, similar to that found by Danks et al.
(1984).  The CN radical is highly density sensitive in the range
of column densities studied here (Federman, Danks, \& Lambert 1984).
We also see a strong correlation between CN and molecular fraction
(\S\ 4.2), and we would expect the latter quantity to also be
correlated with density.  The chemical models of van Dishoeck \&
Black (1988, 1989) appear to trace only the upper envelope of the observed
CN abundances, but this model is for $n$ $\sim$ 500 cm$^{-3}$.
Low- and high-density models of Federman et al. (1984) appear to
bracket our new data as they did for the Federman et al. dataset.

Finally, we see that CO and H$_2$ are also highly correlated.
The slope of the relationship appears to become more gradual at
the highest column densities.  At log $N$(H$_2$) $\sim$ 20.5, the
CO column density begins to increase more rapidly than H$_2$.
However, this is also the point where saturation effects become very
important in assessing $N$(CO) and in several cases a small $b$-value
($\sim$1 km s$^{-1}$) has been assumed which may not be
appropriate if multiple components exist.  Even a modest increase in
the $b$-value can result in a decrease in column density of an
order of magnitude if the line lies on the flat part of the curve
of growth.

On the other hand, chemical models do predict a rapid increase
in CO column density relative to $N$(H$_2$) within the range of
column densities studied in the present work.  The models from van
Dishoeck
\& Black (1988, 1989) show good agreement with the limited data.
Absorption-line measurements of CO are difficult within
the {\it FUSE} bandpass as the rotational structure is poorly
resolved and the lines are often saturated.   {\it HST} observations
of the A--X series of CO lines will be crucial for extending the
CO/H$_2$ ratio.

We note that while the correlations between the line-of-sight
quantities are generally strong, some caution is necessary.  As
we discuss in \S\ 5.2, we can not assess the true distribution
of the majority of the H$_2$ along the line of sight.  Thus, in
cases where the various molecules are distributed across a range
of velocities, the line-of-sight correlations may not be physically
meaningful.  In particular, the column density of CN is sensitive to
particle density and is expected to only trace the densest cloud cores.

\subsection{Total hydrogen column density}
First, we can look at the relationship between the total hydrogen
column density, $N$(H$_{\rm tot}$) = 2$N$(H$_2$) + $N$(\ion{H}{1}), and
color excess; i.e., the gas-to-dust ratio.  Bohlin et al.
(1978) found a linear relationship from {\it Copernicus} data,
$N$(H$_{\rm tot}$) = (5.8 $\times$ 10$^{21}$ cm$^{-2}$) mag$^{-1}$
$E(B-V)$.  In Figure 4, we have plotted the {\it Copernicus/IUE}
dataset, along with our present
{\it FUSE} sample.  The new data fit the old relationship
remarkably well, and the {\it FUSE} data alone give a slope of
5.6 $\times$ 10$^{21}$ cm$^{-2}$ mag$^{-1}$.  There are a few
disagreements larger than the
error bars, but the largest deviation for $E(B-V)$ $>$ 0.3 is for
the {\it Copernicus/IUE} observations of $\rho$ Oph A ($E(B-V)$ = 0.47,
log $N$(\ion{H}{1}) = 21.7).  This deviation
still occurs with the revised {\it IUE} \ion{H}{1} measurement
(Diplas \& Savage 1994)
even though it represents a significant downward revision of the
original {\it Copernicus} value.

This enhanced gas-to-dust ratio
for $\rho$ Oph A has been interpreted as due to a preponderance
of large grains within the $\rho$ Oph cloud (Bohlin
et al. 1978).  These large grains are less efficient at producing
visual reddening, and thus the $E(B-V)$ color excess underestimates
the actual quantity of dust.  The unusual dust properties also
lead to an unusual extinction curve and a very large value of $R_V$.
We note that we have only a single line of sight in our present
sample with $R_V$ $>$ 4 (HD 102065) and we do not have an
independent measurement of $N$(\ion{H}{1}) for this target.

\subsection{Kinetic temperature}
In deriving the temperature of the gas, $T_{01}$, we
assume that the density and column density are high enough such that
thermal proton collisions dominate over other processes in determining
the ratio $N$(1)/$N$(0), and that the observed populations obey the
Boltzmann relation.  In textual form, we will refer to this temperature
as the ``kinetic temperature'', while symbolically we will use $T_{01}$
to emphasize the source of this temperature.  In addition, we emphasize
that this temperature is a ``column-averaged'' temperature, while
the actual temperature will vary throughout the cloud(s) in the line of
sight.

With a ratio of statistical weights, $g_1$/$g_0$ = 9, the
population ratio is simply

\begin{equation}
\frac{N(1)}{N(0)} = 9e^{-E_{01}/kT_{01}},
\end{equation}

\noindent where $E_{01}$/$k$ = 171 K.
With column densities expressed as base-10 logarithms (as
in Table 8), the kinetic temperature (in K) can then be written

\begin{equation}
T_{01} = \frac{74}{\log N(0) - \log N(1) + 0.954}.
\end{equation}

In calculating the uncertainties in kinetic temperature, we take the
combination of 1-$\sigma$
errors that gives the {\it largest} deviation from the best value,
such that the errors on the derived values are more conservative.
Furthermore, in deriving these errors, we have taken 0.04 dex
as the minimum possible error on a column density, even when
we have derived a smaller error.  This corresponds to 10\% and
while this choice is arbitrary we feel that it is a reasonable guess
for the magnitude of any systematic effects.

The average kinetic temperature derived from {\it Copernicus}
observations of 61 lines of sight with log $N$(H$_2$) $>$ 18.0
was 77 $\pm$ 17 K (Savage et al. 1977).  A similar calculation
for the 9 {\it Copernicus} lines of sight with log $N$(H$_2$)
$>$ 20.4, comparable
to the present survey, gives 55 $\pm$ 8 K.  Our {\it FUSE}
sample gives an intermediate value, 68 $\pm$ 15 K.  However,
we note that our sample has a somewhat unusual distribution,
with three lines of sight having $T_{01}$ $\ge$ 94 K,
but none in the range 75--93 K.  In any case, our average
value is similar to that found previously for lines of
sight where H$_2$ is self-shielded.

Despite extending the range of color excess by a factor of 2,
Figure 5 shows that the kinetic temperature in our sample does
not change with increasing $E(B-V)$.  We have also searched for
a correlation between $T_{01}$ and $R_V$ and found none.
We might expect the kinetic temperature to be anti-correlated
with density indicators, and we see such a relationship between
$T_{01}$ and $N$(CN) (Figure 6).  The slope of the relationship is quite
small and there are a few outlying points, but given the small
range in the observed temperatures, the relationship is quite
good.

\subsection{Molecular fraction}

The hydrogen molecular fraction, $f_{\rm H2}$, gives the fraction
of hydrogen atoms in molecular form.  In terms of the column densities
of \ion{H}{1} and H$_2$,

\begin{equation}
{f_{\rm H2}} = \frac{2N(H_2)}{2N(H_2) + N(H I)}
\end{equation}

\noindent The tabulated uncertainties for molecular fraction follow the
same procedure described in the previous section.

The {\it Copernicus} data showed an interesting trend of molecular
fraction with increasing color excess (Savage et al. 1977).
Below $E(B-V) \approx 0.08$, the molecular fraction is quite small,
typically less than 10$^{-4}$, while above this point the fraction
is generally greater than 10$^{-2}$, with few points lying in
between.  This abrupt boundary occurs due to increased self-shielding
of H$_2$ near $N$(H$_2$) $\sim$ 10$^{16}$ cm$^{-2}$, corresponding to
$f_{\rm H2}$ $\sim$ 10$^{-5}$.

In Figure 7, we show molecular fraction versus color excess for
the {\it Copernicus} data and our {\it FUSE} data.  The boundary
at $E(B-V) \approx 0.08$ is not visible because we have chosen a
linear scale for the ordinate.  The {\it FUSE} data mostly overlap
values found previously, but we have greatly increased the number
of lines of sight with at least moderately high molecular fraction.
Even in the range of overlap of the two samples, the {\it FUSE}
sample shows larger molecular fractions, but this is probably a
selection effect.  We do not see an increase in molecular fraction
with increasing extinction within the {\it FUSE} sample.

Figure 8 shows the molecular fraction versus the total-to-selective
extinction ratio, $R_V$, for the {\it FUSE} dataset.  Previous results
have suggested an anti-correlation between the two quantities for
diffuse clouds, consistent with idea that grain coagulation
reduces the available surface area for H$_2$ formation at larger
$R_V$ (Cardelli 1988).  However, our data do not show a statistically
significant relationship between the two quantities.  We do not
presently have good coverage of large values of $R_V$, but several
lines of sight with $R_V$ $>$ 4 remain to be observed as part of
the {\it FUSE} translucent cloud program.

Although little-mentioned, the {\it Copernicus} data show
a good correlation between molecular fraction and kinetic temperature.
Figure 9 shows both data sets, and while the {\it FUSE} data 
themselves show only a weak relationship, those points still follow
the general trend.  There are lines of sight with small molecular
fraction at all kinetic temperatures, but lines of sight with
large molecular fraction are preferentially associated with small
kinetic temperature.  This relationship is not surprising as both
large molecular fraction and small kinetic temperature should be
associated with denser cloud cores.
In a similar sense, we also see a correlation between molecular
fraction and the density-sensitive CN abundance (Figure 10).

\subsection{Extinction curve parameters}
With six extinction curve parameters and four column-density related
quantities ($f_{\rm H2}$, $N$(CH)/$N$(H$_{\rm tot}$),
$N$(CH$^{+}$)/$N$(H$_{\rm tot}$),
$N$(CN)/$N$(H$_{\rm tot}$)), we have 24 potential correlations.
In addition to our {\it FUSE} data points, we have included the
handful of points from the Fitzpatrick \& Massa (1986, 1988, 1990)
and Jenniskens \& Greenburg (1993) extinction curve surveys for which
we have the ancillary data.
We have evaluated the Spearman rank correlation coefficient for
each of these relationships.  In 11 cases the correlation is not
significant at the 1$\sigma$ level, while 10 correlations are
significant at the 2$\sigma$ level.  Thus, we generally either find
no correlation or a good correlation in the statistical sense.
The 2$\sigma$ group includes $N$(CN)/$N$(H$_{\rm tot}$)) vs. both
$c_1$ and $c_2$; $N$(CH$^{+}$)/$N$(H$_{\rm tot}$) vs. $c_3$;
$f_{\rm H2}$, $N$(CH)/$N$(H$_{\rm tot}$), and
$N$(CN)/$N$(H$_{\rm tot}$)) vs. both $c_4$ and $\gamma$; and 
$N$(CH)/$N$(H$_{\rm tot}$) vs. $\lambda_0^{-1}$.
Since the main focus of this paper is H$_2$, we will consider the
two strong correlations involving $f_{\rm H2}$ in detail.

The strongest (3.7$\sigma$) and most intriguing correlation is that
between molecular fraction and the width of the 2175 \AA\ bump, $\gamma$
(Figure 11).
In fact, of all the parameters we have considered in the present
work, $\gamma$ appears to be the best predictor of molecular fraction.
The larger molecular fractions appear to be associated with regions
of larger density.  Thus, our findings suggest that the width of the
2175 \AA\ bump is closely related to
density.  Fitzpatrick \& Massa (1986) reported a similar finding
in a qualitative sense; dense quiescent regions such as dark clouds
and reflection nebulae were associated with broad bumps, and diffuse
clouds and star-forming regions were associated with narrower bumps.
They found a good correlation between $\gamma$ and $E(B-V)$/$r$,
where $r$ is the distance to the star, even
with the biases associated with this density indicator.
Our observed correlation between $f_{\rm H2}$
and $\gamma$ shows much less scatter and could well be taken as linear.
The single outlying point, $\zeta$ Oph, showed the
largest difference in $\gamma$ between the two methods used
by Fitzpatrick \& Massa (1986, 1990).  We have used the final value
of Fitzpatrick \& Massa (1990)
based on the overall extinction curve fits, but the initial value,
based on just the region around the bump, would lie much closer to
the rest of the points.  The authors attributed this large
difference to the relative shallowness of the bump and the
resultant uncertainty in separating the bump from the rest of
the extinction curve.

Ignoring $\zeta$ Oph, an unweighted linear fit of the rest of
the points gives the relation,

\begin{equation}
f_{\rm H2} = 1.44\gamma - 1.04
\end{equation}

\noindent corresponding to a minimum value of $\gamma$ of 0.72, and
a maximum value of 1.42.
(If we include $\zeta$ Oph in the fit, the allowed range is 0.67--1.51.)
In fact, the extrema in $\gamma$ in the
entire Fitzpatrick \& Massa (1990) and Jenniskens \& Greenberg (1993)
samples of more than 100 curves are 0.76 and 1.383 (or 1.25 if we
ignore $\zeta$ Oph).

While the bump width appears to be a good predictor of molecular
fraction in our sample, this relationship may not hold in all
environments.  For example, many lines of sight in the SMC have
no discernible 2175 \AA\ bump at all (Gordon \& Clayton 1998).  On
the other hand, in a survey of 30 Galactic lines of sight selected
to sample low-density gas, Clayton, Gordon, \& Wolff (2000)
found small values of $\gamma$ ($\lesssim$1.0).  This finding
is consistent with the likelihood that the lines of sight sample
gas with low molecular content.  Thus, a strong correlation
between molecular fraction and bump width may apply in most
Galactic environments, albeit the relationship may not be linear.

The strength of the far-UV curvature, $c_4$, exhibits the other
strong correlation with $f_{\rm H2}$, at the 2.6$\sigma$ level
(Figure 12).  The extrema in $c_4$ are HD 102065 and HD 62542, and
these points show up as outliers in the relationship.
Although neither line of sight has a well-determined value
of $f_{\rm H2}$ it is highly unlikely that our reported values are
so inaccurate as to match the apparently linear trend with $c_4$ seen
in the other points.  The presence of this correlation is consistent
with the previously noted tendency for a steep far-UV rise to be
associated with a broad 2175 \AA\ and the observed correlation between
the bump width and molecular fraction.

An alternative explanation for the correlations between molecular
fraction and both the bump width and far-UV curvature concerns
the properties of the dust grains.  Increased far-UV curvature
is thought to be associated with smaller than normal dust grains
(Cardelli, Clayton, \& Mathias 1989).  The 2175 \AA\ bump is
most likely associated with small carbonaceous grains (D\'{e}sert,
Boulanger, \& Puget 1990), and perhaps smaller grains lead to
broader bumps.

Grain size is also thought to be smaller in lines of sight with
small $R_V$, and Cardelli (1988) found an inverse correlation
between molecular abundances and $R_V$.  He attributed this
correlation to the effects of these smaller grains and their
effect on H$_2$ formation and destruction.  With similar total
grain masses, the smaller grains will provide greater surface area,
yielding a greater H$_2$ formation rate, and a smaller
photodissociation rate via the increased far-UV extinction.
We recall, however, that in the present work, we do not find
a good correlation between molecular fraction and $R_V$.

\section{Discussion}
In this section, we will mainly focus on what our findings say
about the nature of our present lines of sight relative to diffuse
clouds.  The overall line-of-sight characteristics
of most of our present sample satisfy the criterion to be considered
``translucent'', i.e. $A_V$ $\gtrsim$ 1.  Implicit in the definition
of a translucent cloud is that we are considering a single molecular
cloud, and not a collection of several diffuse clouds.
For the purposes of this discussion, we adopt a definition of
a ``translucent cloud'' similar to that envisioned by van Dishoeck \&
Black (1988); i.e. $f_{\rm H2}$ $\gtrsim$ 0.9, $T_{01}$ $\lesssim$
40 K, and $A_V$ $\gtrsim$ 1.  Such a cloud may be an isolated cloud,
a skin around a dense cloud, or a core located within significant
diffuse material.

If a line of sight is dominated by one of these clouds, we would
expect this situation to be reflected in several of our measured
quantities.  Specifically, the observed molecular fraction should
be large, while the kinetic temperature should be small.  As shown
in Figure 9, these two quantities do indeed show an anti-correlation,
with considerable scatter.  Despite the scatter, all of the lines
of sight where $N$(H$_2$) $>$ $N$(\ion{H}{1}) ($f_{\rm H2}$ $>$ 2/3)
show small kinetic temperatures.  In fact, there appears to be
a distinct group of 10 lines of sight centered near $T_{01}$ = 55 K and
$f_{\rm H2}$=0.7 that is
separated from the rest of the sample\footnote{These 10 lines of sight are:
HD 24534, HD 27778, HD 62542, HD 73882, HD 99675, HD 154368, HD 210121,
and the {\it Copernicus} targets $\zeta$ Oph, $o$ Per, and $\zeta$ Per}.
This is even more apparent if we ignore the {\it FUSE} data points
without direct measurements of $N$(\ion{H}{1}).

However, there are several lines of sight with similar or even
smaller kinetic temperatures than this group of 10.  In addition,
while the molecular fractions are relatively large, they do not
closely approach unity as we might expect.  Also, the extinctions
for several of these lines of sight are less than or equal to
one magnitude, barely satisfying the rather loose definition of
a translucent cloud we have adopted.  To further assess the question
of whether we are seeing individual translucent
clouds, we need to consider the distribution of material along the
line of sight (\S\ 5.1).  We also consider evidence from
studies of chemical depletions (\S\ 5.2).  Finally, we consider
the question of why we see few, if any, translucent clouds in our
lines of sight (\S\ 5.3).

\subsection{Multiple clouds and ``hidden'' translucent clouds}
Arguing in favor of the hypothesis that we are seeing at least
a few translucent clouds is the fact that the overall line-of-sight
column densities can be greatly affected by the particular
distribution of material.  Even if highly molecular material
exists, there could be a skin of diffuse material surrounding
this cloud, or additional diffuse clouds along the line of
sight.  In these cases, the observed integrated molecular fraction
could be considerably less than unity and the kinetic temperature
could be affected as well.  Thus, even a line of sight with only a
moderately high molecular fraction could harbor a translucent
cloud.

Chemical modeling of the lines of sight can, in principle, help
constrain the distribution of material.  Models have been
constructed which reproduce the column densities of a variety
of species in diffuse clouds, including H$_2$ (e.g., van Dishoeck
\& Black 1986).  In some lines of sight, the models have difficulty
reproducing the high-$J$ column densities of H$_2$ (e.g., Paper II).

Part of our group has developed a new code to study the formation,
destruction, radiative transfer, and ro-vibrational excitation
of H$_2$ (Browning et al. 2002).
Several lines of sight with published high-$J$ column densities
for H$_2$ have been modeled with this code, including HD 110432
and HD 73882.  In many cases, it was difficult to model a line
of sight as a single cloud without invoking extreme conditions,
such as a radiation field 10--100 times the Galactic mean, which
is unrealistic in most cases.  Better matches require multiple
clouds and/or multiple pathways for incoming UV radiation.
In addition, changes in one physical parameter can
mask changes in another such that a group of models corresponding to
a range of physical conditions will all match the column densities
for an individual line of sight.  Ensembles of models,
combined with large samples of lines of sight, can provide more
definitive results, such as in the LMC/SMC (Tumlinson et al. 2002).

We can also observationally study the possibility that the H$_2$
along these
lines of sight lies in several distinct clouds, with corresponding
smaller extinctions for each cloud.  The very high resolution
ground-based data show that most lines of sight have numerous
components in \ion{Na}{1} and \ion{K}{1}, and the strongest
components usually lie within a 10 km s$^{-1}$ velocity range
(D. E. Welty et al. 2002, in preparation).  Data for the CH
radical show similar behavior.  Given the good correlations
between $N$(H$_2$) and these species, we might expect the H$_2$
to also be distributed among several components.

We can assess the distribution of H$_2$ indirectly, in
several ways.  Equivalent width data allow us to perform a
curve-of-growth analysis of the high-$J$ lines.  Most simply, we
can construct a best-fit single-component curve of growth via
fitting techniques and determine the best values for the column
densities of each $J$-level as well as the optimum ``effective''
$b$-value of the distribution.  However, we can not always assume
that the actual component structure is well approximated by a
single $b$-value.
Our program of high-resolution spectroscopy allows us to better
constrain the component structure with the assumption that some or
all of the components seen in \ion{Na}{1} and/or \ion{K}{1} and/or
CH contain H$_2$, with perhaps a scaling of the $b$-values of the
individual components, and possible variation in the ratios of
the ground-based species to H$_2$ from component to component.
As we have seen, for the overall lines of sight, there is a nearly
one-to-one correspondence between CH and H$_2$, and strong
correlations also appear for \ion{Na}{1} and \ion{K}{1} (Welty
\& Hobbs 2001).

In our preliminary investigations of the high-$J$ H$_2$ lines, we
often find a better match with ``effective'' $b$-values greater
than can be obtained with the CH component structure.  We would
then have to use additional components observed in \ion{K}{1} or
\ion{Na}{1} --- including some of the weaker
components, as compared with only using the strongest.  Thus, even
with a ``simple'' component structure, the H$_2$ still appears
to be distributed among several closely-spaced velocity components.
Despite the small separations, the components are resolved in the very
high resolution optical spectra, which suggests discrete components and
not just material more-or-less uniformly distributed in velocity space.
With the additional components that may contribute at least small
amounts of H$_2$ to the overall observed column densities, the
quantity of H$_2$ available in each component is further diluted.

One important caveat to this discussion of the component structure
is that in looking at the high-$J$ lines we are considering only
1\% or less of the total H$_2$.  We have little hope of directly
assessing the true component structure of the extremely strong
$J$ = 0--1 lines.  In addition, the $J$ = 2 lines are also very strong
and heavily saturated or even damped, and will not always provide
constraints to the component
structure.  We expect the material containing the low-$J$ lines
to mostly be located in the self-shielded cloud core(s), while the
material containing the high-$J$ lines may be more physically
widespread along the line of sight.  Thus, the component structure
corresponding to $J$ = 0 and 1 may be simpler than that for
the high-$J$ lines.

There is a tendency for the lines of sight with the largest
molecular fractions to have the simplest component structures when
looking at CH or the strong \ion{Na}{1} and \ion{K}{1} components
that are most likely to also contain H$_2$ (D. E. Welty et al. 2002,
in preparation).  In Table 10 we give preliminary information on the
CH component structures observed at high resolution.  We caution
that the 4 cases with a single component are for observations at
the lowest resolutions.  The typical spacing between components
is 3--5 km s$^{-1}$, so there may be unresolved structure, as
indicated in higher resolution \ion{K}{1} spectra in these 4 cases.
In fact, we find more complex structure in the \ion{K}{1} spectra in
all lines of sight in Table 10.  It is more difficult to detect weak
components or resolve closely spaced components in CH observations
due to $\Lambda$-doubling, the small atomic weight, and the relatively
weak spectral lines.  In the CH data with resolution of less than
2 km s$^{-1}$, the lines of sight with more material in the strongest
component and/or fewer components have larger molecular fractions.
These lines of sight presumably have less contamination from
foreground clouds of atomic gas.
Yet, in all cases where we have this ``very high'' resolution
data, there are multiple components, with no more than $\sim$75\%
of the molecules in the dominant component.

Given the relatively close spacing of the CH components we have to
consider whether they represent small knots of material within
a single cloud, or distinct clouds along the line of sight.
If we are dealing with knots of material within a single cloud,
we still might consider this cloud to be a translucent cloud due
to the large overall extinction and high molecular content.  Also,
within a single cloud, the extreme widths of the H$_2$ $J$ = 0 and
1 lines effectively self-shield the entire range of velocities.
If we are dealing with multiple distinct clouds at the even closer
spacing suggested by the \ion{K}{1} data, the amount of molecular
material within the most abundant cloud core will generally be even
less than indicated in Table 10 (or in our mathematical analysis below
and in Table 11).  This suggests that these lines of sight are simply
collections of many diffuse clouds with small extinction, albeit
with large molecular fractions and presumably relatively high
densities.

An indirect method for assessing the quantity of H$_2$ and
\ion{H}{1} in each component comes from the study of chlorine.
The chemistry of chlorine is intimately connected to that of
H$_2$ because of the large reaction rate with \ion{Cl}{2}.
With a first ionization potential of 13.01 eV, chlorine will be
primarily ionized in \ion{H}{1} gas,
and primarily neutral in association with H$_2$ (Jura 1974;
Jura \& York 1978).  Thus, observations of \ion{Cl}{1} and
\ion{Cl}{2} lines (in the UV) can constrain the hydrogen
molecular fraction.  P. Sonnetrucker et al. (2002, in preparation)
have used this technique to infer that the true molecular fraction
in the dominant area of H$_2$ toward HD 192639 is as large as twice
the line-of-sight value (2/3 vs. 1/3).  High-resolution
measurements of the the \ion{Cl}{1} $\lambda$1347 line may provide
important constraints on the molecular fractions of individual
velocity components toward our lines of sight.

While detailed physical and chemical modeling of these lines
of sight will appear in a later paper or papers, we can perform
a feasibility analysis on the H$_2$ column densities in each line
of sight to assess the possible presence of translucent clouds.
I.e., we know the overall line of sight parameters, and we also
know approximately what a translucent cloud should look like.
Thus, we can add the expected observed properties of putative
translucent clouds to diffuse cloud material in various
combinations that produce the observed total line of sight quantities.
While not a physical model of the clouds, it does provide interesting
constraints on the observable quantities.

We define translucent material as having $f_{\rm H2}$ = 0.9 and
$T_{01}$ = 30 K, comparable to the parameters used in the translucent
cloud models of van Dishoeck \& Black (1988).  
To constrain the total column density of the translucent cloud
and the remaining material, we assume that the previously discussed
relationship between $E(B-V)$ and $N$(H$_{\rm tot}$) still applies
for highly molecular clouds.  We then mathematically create the
largest embedded translucent cloud that is still consistent with the
line of sight totals, and that produces realistic parameters for
the remaining ``diffuse'' material, i.e. $T_{01}$ $<$ 200 K.
If the line of sight molecular fraction is
small and the temperature is large, only a small amount of material
could possibly be locked up in a cold, highly molecular cloud, and
vice versa.

In Table 11 we give our results for this breakdown of material
into translucent and diffuse clouds.  From this simulation,
we see little evidence for individual translucent
clouds along our lines of sight.  In two directions, HD 73882 and HD
154368, translucent clouds with $A_V$ $\sim$ 1 could exist.  The
remaining diffuse material still has a molecular fraction appropriate
for the color excess.  HD 62542 is the only direction where more than
half of the material could be translucent.

If we were to relax the criteria for a ``translucent'' cloud to
something like $f_{\rm H2}$ = 0.8 and $T_{01}$ = 40 K, comparable
to the most extreme line of sight values we actually see, we would
then have many more cases where there is as much translucent material
as diffuse.  Not coincidentally, this larger list of lines of sight
with as least as much translucent material as diffuse is identical
to the list of {\it FUSE} targets in the upper left-hand portion
of Figure 9 discussed at the beginning of \S\ 5.  However, we would
still not have additional translucent clouds with $A_V$ $>$ 1.

We have assumed a gas-to-dust ratio of 5.8 $\times$ 10$^{21}$ cm
$^{-2}$ mag$^{-1}$ based on the excellent match we see in Figure 4.
The observed scatter could support variations on either side of
this relationship for our {\it FUSE} targets, albeit not as large
as seen for $\rho$ Oph.
However, we would need large (and unrealistic) differences in the
dust characteristics between the ``diffuse'' and ``translucent''
material to modify our conclusions in the previous paragraphs.

\subsection{Evidence from chemical depletions}
As noted in \S\ 2.3.13, chemical depletions can be used as an
indicator of the presence of translucent clouds.  For HD 192639,
the depletions of about a dozen species indicate similar conditions
to diffuse clouds (P. Sonnentrucker et al. 2002, in preparation).
We have also
undertaken a study of \ion{Fe}{2} depletions in most of our {\it FUSE}
targets (Snow, Rachford, \& Figoski 2002).  Snow et al. found
that the depletion of iron is more or less uniform for the {\it FUSE} lines
of sight.  There is little evidence for increased depletion with
increasing extinction, molecular fraction, or $N$(H$_{\rm tot}$)/$r$,
within the range of $E(B-V)$ and $A_V$ covered by {\it FUSE}.  The
new observations did not extend the trends of increasing depletion
with increasing density found by Savage \& Bohlin (1979) and
Jenkins, Savage, \& Spitzer (1986).  Such increased depletions are
expected to occur at some point within the translucent clouds
(Snow et al. 1998).  Given the moderate resolution of the {\it FUSE}
\ion{Fe}{2} observations, we can not rule out the possibility of large
depletions for individual components, which become ``smoothed out'' in
the line-of-sight values.

Two of our {\it FUSE} targets with particularly large molecular fraction,
HD 24534 and HD 154368, have been observed at high resolution with
{\it HST} to search for increased depletions in several additional
species.  The results for HD 24534 show a very slight, and probably
not statistically significant, increase in carbon depletion relative
to diffuse clouds (Sofia, Fitzpatrick, \& Meyer 1998).  Snow et al.
(1998) did not find increased oxygen depletion toward HD 24534.  Snow
et al. (1996) studied many species toward HD 154368 and also did not
find increased depletions.  They concluded that the line of sight toward
HD 154368 contained extended regions of moderate-density gas instead of
one or more dense cloud cores, despite $A_V$ $\approx$ 2.5.

\subsection{Where are the translucent clouds?}
While each of techniques we have discussed has limitations, we
are forced to conclude that with few exceptions there is little
evidence for individual translucent clouds within our lines of sight,
based on the definition given at the beginning of this section.
For HD 24534, HD 154368, and HD 192639 we have considerable
information that leads to this conclusion, while the evidence
for the rest of the lines of sight is more circumstantial.
There are several possible reasons why we have not found such
clouds:

1. Translucent clouds are just beyond our {\it FUSE} range. \\
Very few stars have enough UV flux to be observed with {\it FUSE}
through $A_V$ $>$ 2, and our present sample includes only seven such targets.
Many interesting targets were rejected from the original potential
target list for being too faint.  Several of these targets have $A_V$
= 2--4 (e.g. HD 80077, HD 169454) and lie behind molecular clouds.
When combined with the fact that many, if not most, lines of sight will
have contaminating diffuse material, we may need to reach $A_V$ $\sim$
3--5 to find evidence for individual clouds with molecular fractions near
unity and $A_V$ $\sim$ 1--2.

Chemical models suggest that several important transitions involving
carbon take place in clouds in the range $A_V$ $\sim$ 1--3, such as
increased CO abundance and a dramatically larger
\ion{C}{1}/\ion{C}{2} ratio (e.g., van Dishoeck \& Black 1988, 1989).
There is also evidence for changes in grain conditions at $A_V$ $\sim$ 3.
For example, Whittet et al. (2001) find an increase
in $R_V$ in the Taurus dark clouds for $A_V$ $>$ 3.  The 3-$\mu$m
water-ice feature begins to appear at this point as well,
lending support for the idea that grain mantle growth becomes
important here.

2. $\zeta$ Oph-type lines of sight are indeed ``translucent.'' \\
As we have mentioned, large molecular fractions do seem to be
correlated with not only simpler component structures, but also
low kinetic temperatures, large column densities of density-sensitive
species such as CN, and the environment-sensitive 2175 \AA\ bump
width and far-UV extinction curvature.  If we were to relax the
requirement for large $A_V$, we could consider all 10 lines of sight
discussed in the preamble to this section as containing ``translucent
clouds.''  However, we note that even within this group, the already
small $A_V$ is sometimes broken up into even smaller clouds (such as
toward $\zeta$ Oph).  We are reluctant to accept this possibility
without a confirmation that we do not see large molecular fractions
in appropriate targets with large $A_V$.

3. There is no distinction between ``diffuse'' and ``translucent''
except larger $A_V$. \\
This possibility is the simplest.  However, there is significant
evidence listed above to support the idea of at least a gradual
transition in physical parameters through $A_V$ = 1--5, if not
more abrupt transitions.  Again, we feel that we need to cover
a larger range of $A_V$ before we could accept this possibility.

We feel that the first possibility is the most likely one based
on our present findings.  The ubiquity of interstellar clouds
makes it difficult to find lines of sight that only sample a
single cloud, or stars that are still observable through an ensemble
of several clouds that might contain a translucent cloud.

A more definitive answer to the question posed in this section
would require H$_2$ observations of several lines of sight with
$A_V$ $>$ 2 where the neutral atoms and simple carbon-containing
molecules show relatively simple velocity structure such as HD
169454 (Jannuzi et al. 1988; Crawford 1997).  Even in this case,
the \ion{Na}{1} component structure is more complicated (Federman
\& Lambert 1992).
A more serious problem is that {\it FUSE} is barely capable of this
observation, and is incapable of observing several other promising
targets such as HD 80077.  We note that while the decrease in 
observed UV flux due to increasing H$_2$ absorption plays a role
in limiting the number of targets we can access, it is a very
small role.  For instance, we can calculate model H$_2$ absorption
spectra and compare the height of the
peaks between the vibrational bandheads for N(H$_2$) = 1 $\times$
10$^{21}$ (present sample) and 5 $\times$ 10$^{21}$ cm$^{-2}$
(a highly molecular cloud with $A_V$ $\approx$ 5--6).
For this extreme example, the intensity relative to the continuum
at the peaks on either side of the (4,0) bandhead is $\approx$0.85
for the low-column case, and $\approx$0.45 for the high-column cases.
However, the effect at the (2,0) and (1,0) bandheads is much smaller
since these bands are weaker, i.e., $\approx$0.95 versus $\approx$0.75.
This is a trivial effect relative to the increased dust extinction,
and indicates that our dataset is not significantly biased against
high-column lines of sight through selection effects.

As a final note, we emphasize that there is a subset of lines of
sight with relatively small extinction, but extreme line of sight
characteristics.  These include HD 62542 and HD 210121, both
of which appear to have a large molecular fraction, and are among
the best candidates for having translucent clouds based on Table
10.  We may eventually find lines of sight with larger $A_V$ that
indeed contain individual clouds similar to those toward
HD 62542 and HD 210121 and thus the latter would be considered
translucent clouds.

\section{Summary}
We have completed the first {\it FUSE} survey of molecular hydrogen
in lines of sight with $A_V$ $\gtrsim$ 1.  The survey includes
observations toward 23 early-type stars, which sample gas in a range
of environments.
Through profile fitting of three vibrational bandheads, we have
directly measured the H$_2$ column densities in the $J$ = 0 and
$J$ = 1 states.  Combined with ancillary data, we have then derived
the total H$_2$ column density, hydrogen molecular fraction, and
kinetic temperature.  In addition, we have compiled a
set of extinction parameters for our lines of sight.

With this information we have investigated many important
relationships between parameters.  We have extended previous
correlations between $N$(H$_2$) and $N$(CH), $N$(CH${^+}$),
$N$(CN), and $N$(CO).  While the formation of CH$^{+}$ is still an open
question, the other correlations closely match predictions from
chemical models, despite possible differences in the distribution
of these species.  We find a potentially powerful combination in
assessing the H$_2$ content of a line of sight
based on ground-based measurements and {\it IUE} or {\it HST} mid-UV
observations.  The excellent relationship between $N$(H$_2$) and $N$(CH)
gives a good estimate of $N$(H$_2$), while the measurement of the width
of the 2175 \AA\ bump gives a good estimate of molecular fraction.

We find a generally self-consistent picture of these lines of sight
in the sense that various indicators of density correlate well with
each other; i.e., molecular fraction, kinetic temperature, CN abundance,
and extinction curve parameters.

While our sample has a relatively large average molecular fraction,
we have not found lines of sight with molecular fractions greater
than 0.8.
We have identified a subset of 7 lines of sight (plus 3 lines
of sight observed by {\it Copernicus}) with large molecular fraction
and small kinetic temperature which represent the best candidates
for the presence of ``translucent cloud'' material.  However, the
possible quantities of material that could be associated with
translucent clouds generally corresponds to $A_V$ $<$ 1 mag, the
nominal lower limit for such clouds, and in no case do we see
evidence for highly molecular material corresponding to $A_V$ $\gtrsim$
1.5 mag.  In addition, in most cases we see evidence for multiple
velocity components for H$_2$, which would further divide up the
extinction when considering the individual clouds.  Thus, our
conclusion is that for the present sample we are observing only
a few individual translucent clouds.  Rather, we are mostly seeing
combinations of diffuse clouds.  We suggest that without specific
evidence to indicate individual translucent clouds, lines of
sight with $A_V$ $\gtrsim$ 1 should be called ``translucent lines of
sight'' and the term ``translucent cloud'' should be avoided.

An ongoing analysis of the high-$J$ lines of H$_2$ in these lines of
sight, combined with a detailed modeling program similar to that for the
Magellanic Clouds (Tumlinson et al. 2002), will allow us to better
assess the physical conditions in the clouds.  This will be the first
large study of high-$J$ excitation of H$_2$ to incorporate very high
resolution ground-based observations of species such as \ion{K}{1},
\ion{Na}{1}, and CH to help assess the unresolved component structure
of H$_2$.  Also, we are undertaking detailed analysis of all available
species in selected individual lines of sight (e.g. P. Sonnentrucker,
et al. 2002, in preparation, for HD 192639), combining {\it FUSE}
data with recently obtained {\it HST} data.  Further studies with
{\it HST} in the lines of sight with the largest molecular fractions
will also be useful.

Finally, several additional lines of sight from the {\it FUSE}
translucent cloud survey have already been observed since the cut-off
date for inclusion in the present work (June 2001), and we anticipate
that most of the 21 remaining targets will be observed.  Several very
important targets such as HD 37903 and HD 147889 are part of the
program.  These additional observations will double the number of
lines of sight studied, and should improve and/or extend the results
of the present work.

\acknowledgments
We thank the anonymous referee for many helpful comments.
This work is based on data obtained for the Guaranteed Time Team by the
NASA-CNES-CSA {\it FUSE} mission operated by the Johns Hopkins University.
Financial support to U.S. participants has been provided by
NASA contract NAS5-32985.  This research has made use of the SIMBAD
database, operated at CDS, Strasbourg, France.

\clearpage
\begin{figure}
\plotone{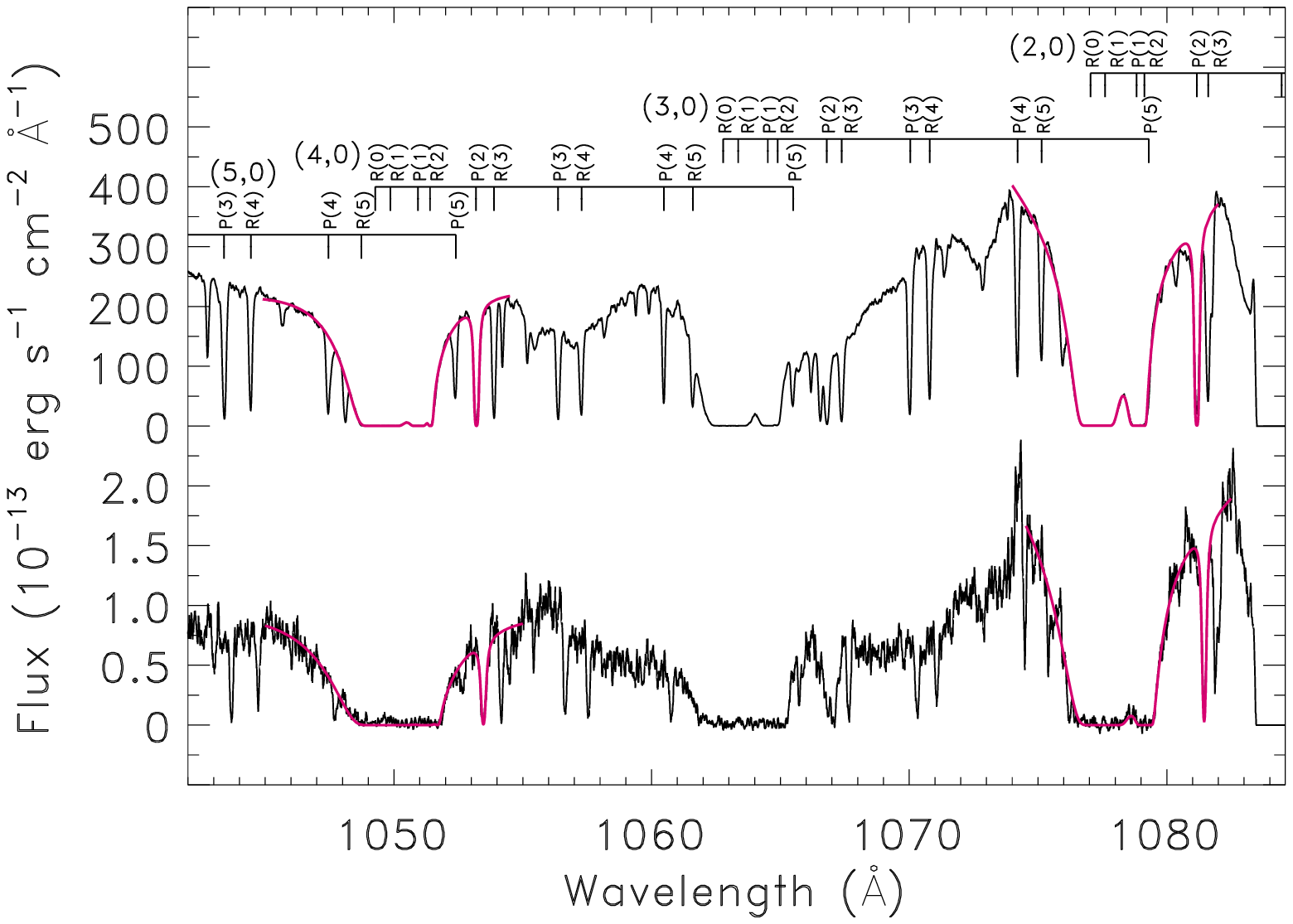}
\caption{Sample spectra at high and low S/N.  Top - HD 210839 (peak S/N
$\approx$ 24 per pixel); bottom - HD 154368 (S/N $\approx$ 2).  Both
spectra have been smoothed to resolution ($\sim$9 pixels) for
presentation.  This spectral region includes the Lyman series (4,0),
(3,0), and (2,0) vibrational bandheads of H$_2$ (from left to right),
as well as broad stellar features at 1057 \AA\ and 1073 \AA .  A less
obvious stellar feature lies near 1067 \AA .  The numerous narrow lines
are mostly due to rotationally excited H$_2$, as indicated by the series
of tickmarks above the spectra; exceptions include \ion{Ar}{1}
$\lambda\lambda$ 1048,1067 and \ion{Fe}{2} $\lambda$1055.  Profile fits
to the (4,0) and (2,0) bands are shown.}
\end{figure}

\clearpage
\begin{figure}
\plotone{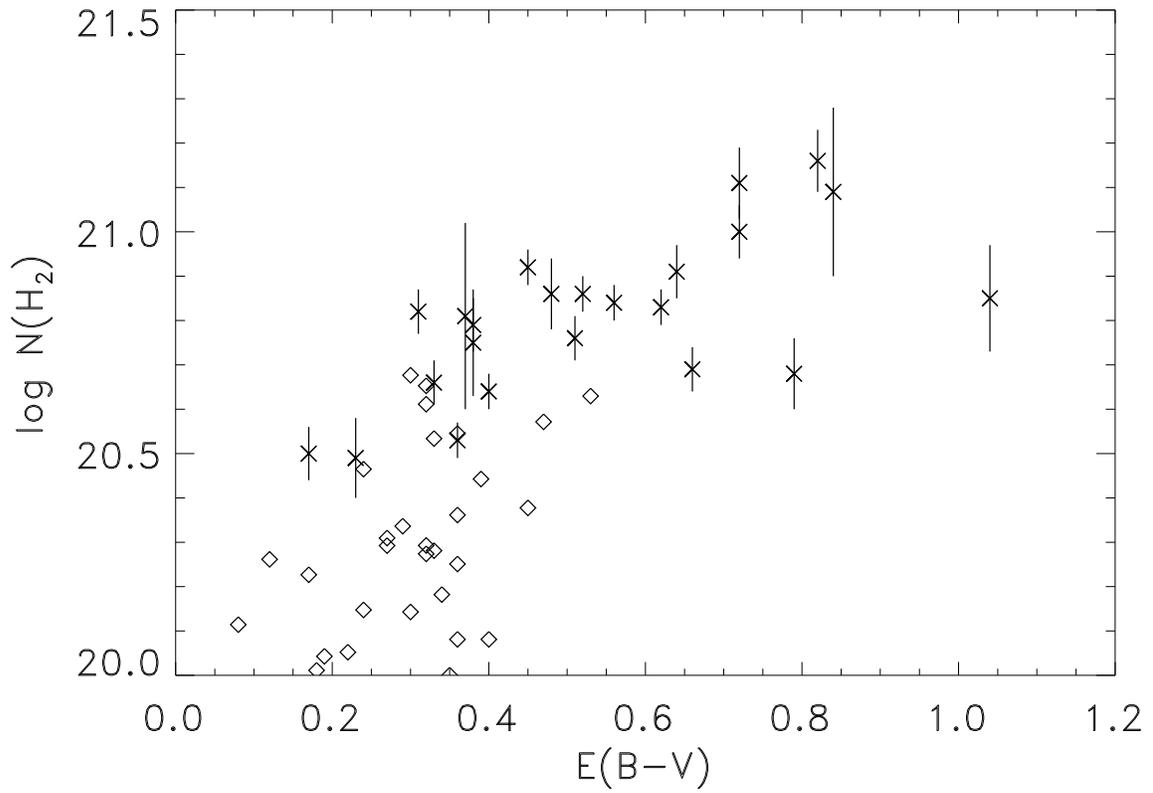}
\caption{Molecular hydrogen column density vs. color excess.
Symbols - crosses: {\it FUSE}; diamonds: {\it Copernicus}}
\end{figure}

\begin{figure}
\plotone{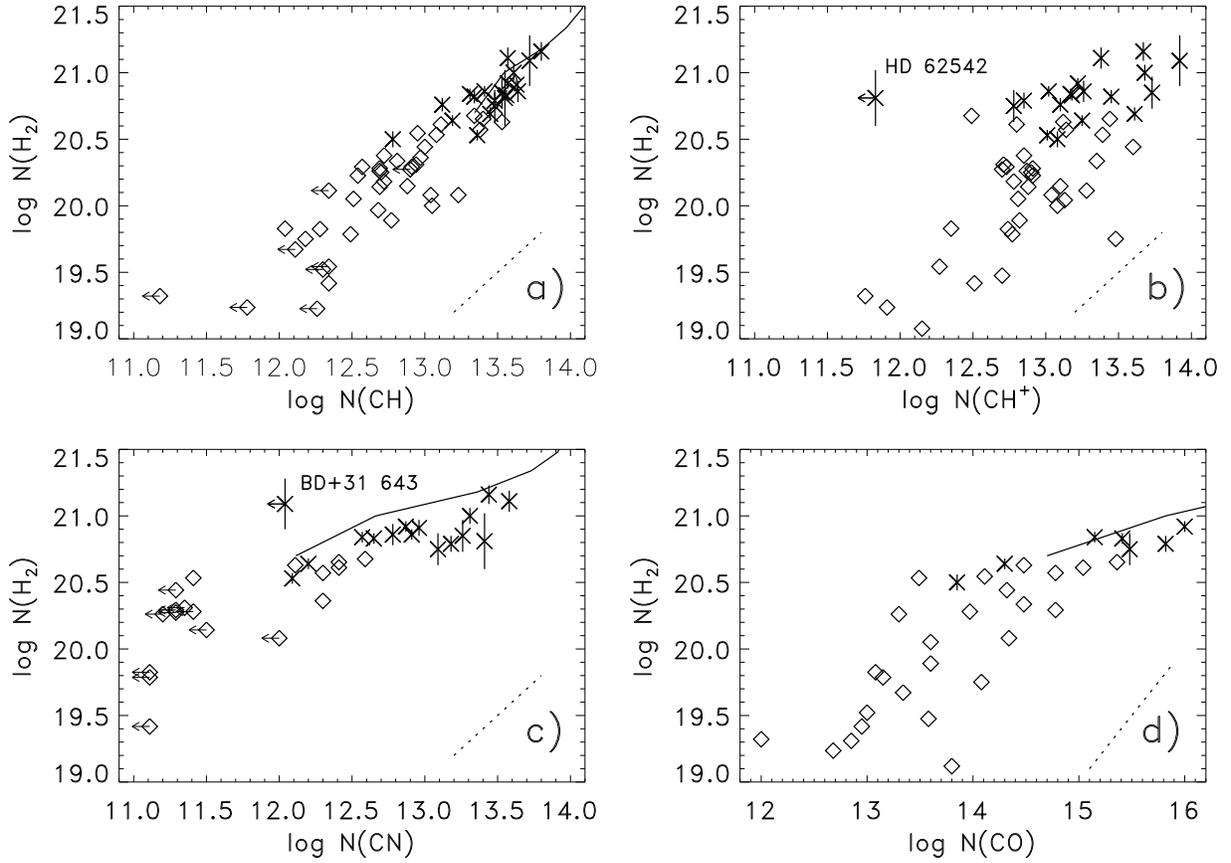}
\caption{H$_2$ column density vs. other molecular column densities.
Symbols as in Figure 2.  Solid curves in panels a), b), and d) are
translucent cloud models from van Dishoeck \& Black (1989).  Dotted
lines in the bottom right corner of each panel correspond to unit
slope for that panel.}
\end{figure}

\clearpage
\begin{figure}
\plotone{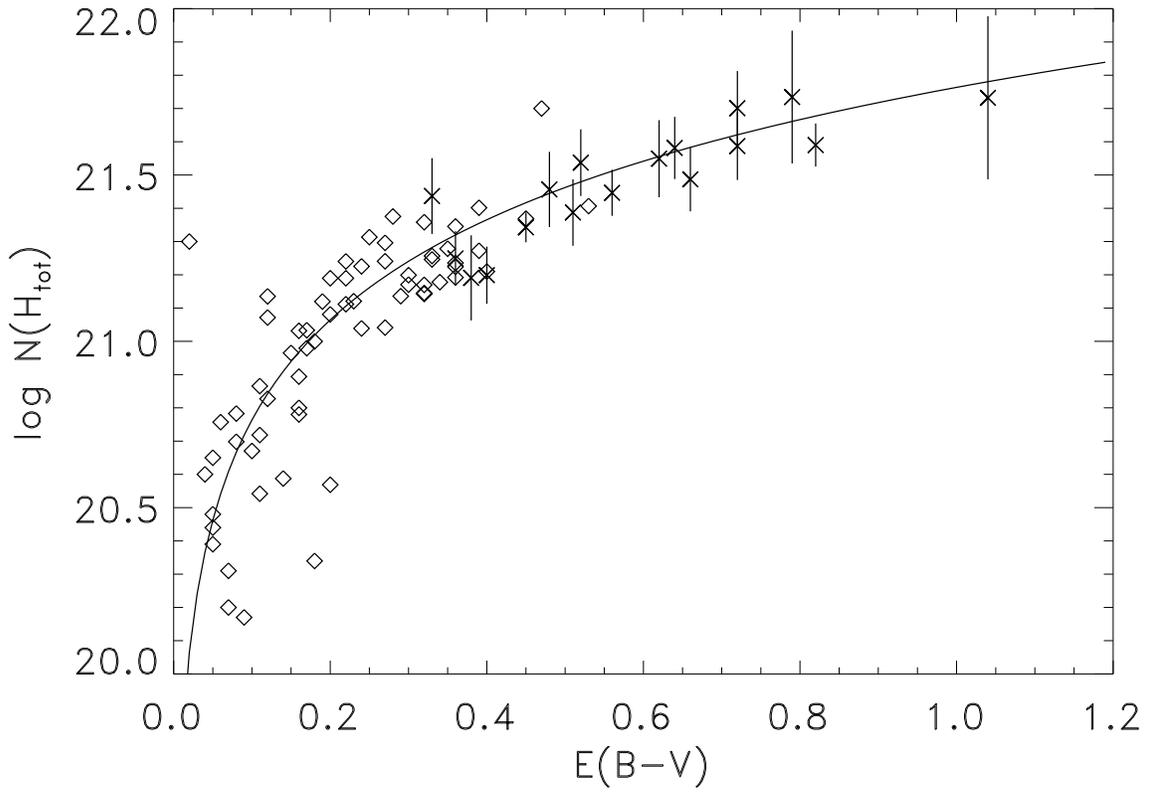}
\caption{Symbols as in
Fig. 2.  The six {\it FUSE} points with no independent measurement
of $N$(H I) are not included.  The solid line corresponds to the
relation $N$(H$_{\rm tot}$) = (5.8 $\times$ 10$^{21}$ cm$^{-2}$
mag$^{-1}$)$E(B-V)$ given by Bohlin et al. (1978).}
\end{figure}

\clearpage
\begin{figure}
\plotone{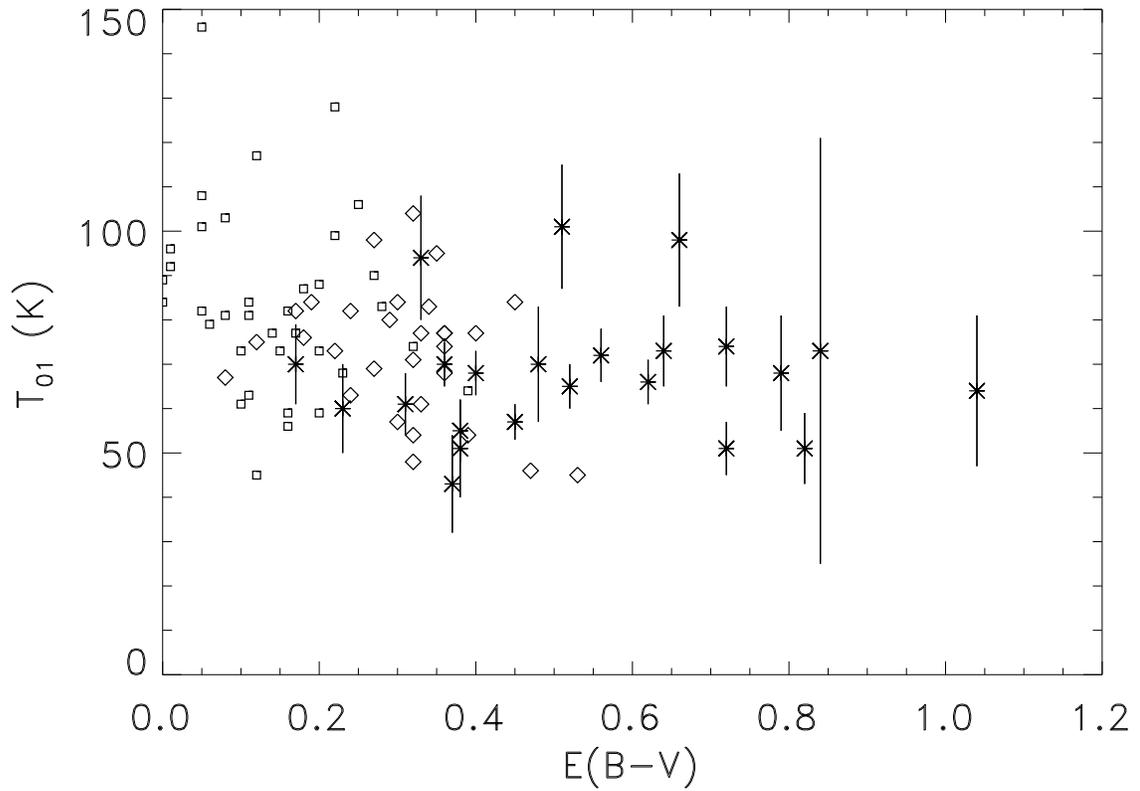}
\caption{Kinetic temperature vs. color excess.  Symbols - crosses: {\it FUSE};
diamonds: {\it Copernicus} points with $N$(H$_2$) $>$ 10$^{20}$
cm$^{-2}$; squares: {\it Copernicus} points with $N$(H$_2$) $<$ 10$^{20}$
cm$^{-2}$.}
\end{figure}

\begin{figure}
\plotone{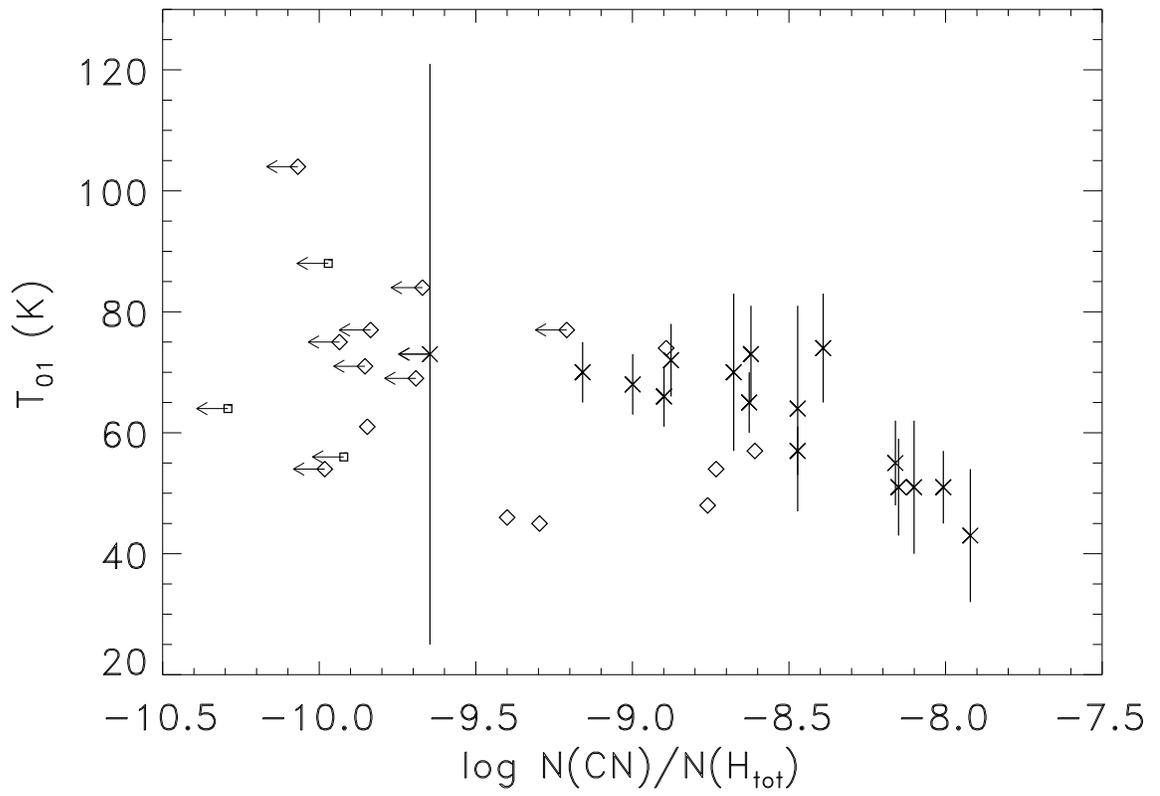}
\caption{Kinetic temperature vs. fractional CN abundance.  Symbols as
in Figure 5.  (Note the particularly large error bar for BD $+$31$^{\circ}$ 643
on this expanded vertical scale.)}
\end{figure}

\clearpage
\begin{figure}
\plotone{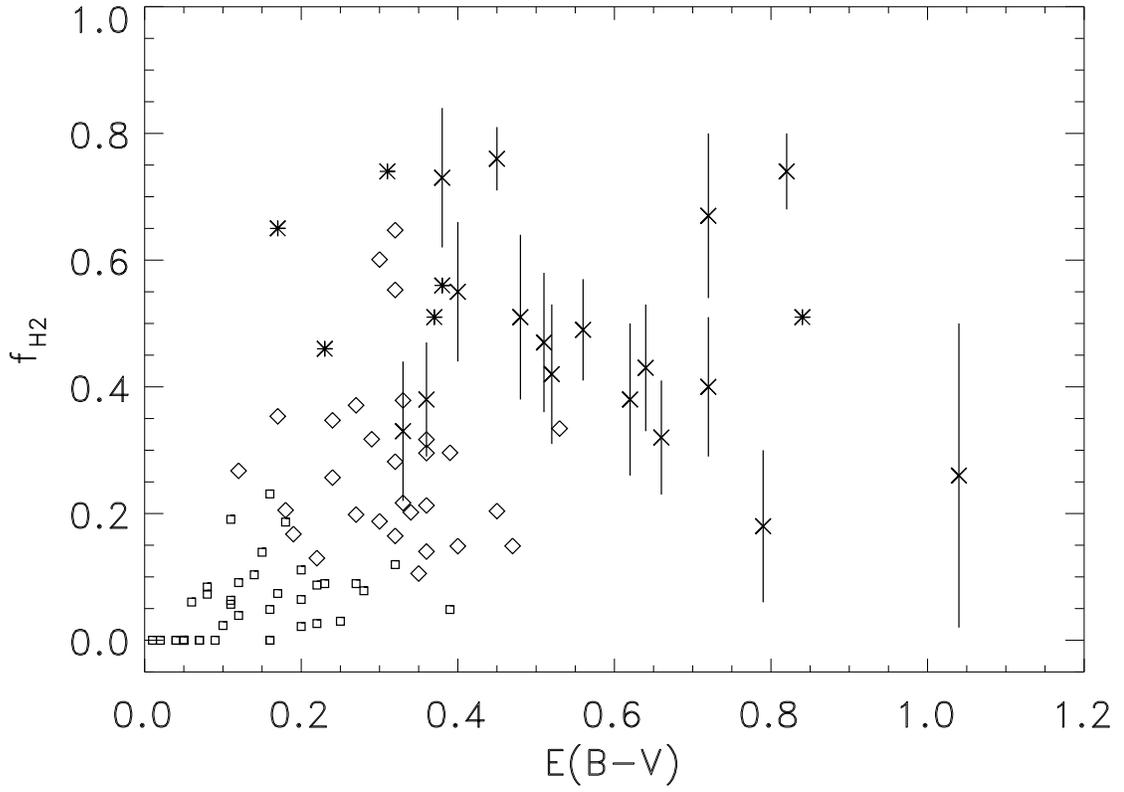}
\caption{Molecular fraction vs. color excess.  Symbols - crosses: {\it FUSE};
asterisks: {\it FUSE} points with no independent measurement of N(H I);
diamonds: {\it Copernicus} points with $N$(H$_2$) $>$ 10$^{20}$
cm$^{-2}$; squares: {\it Copernicus} points with $N$(H$_2$) $<$ 10$^{20}$
cm$^{-2}$.}
\end{figure}

\clearpage
\begin{figure}
\plotone{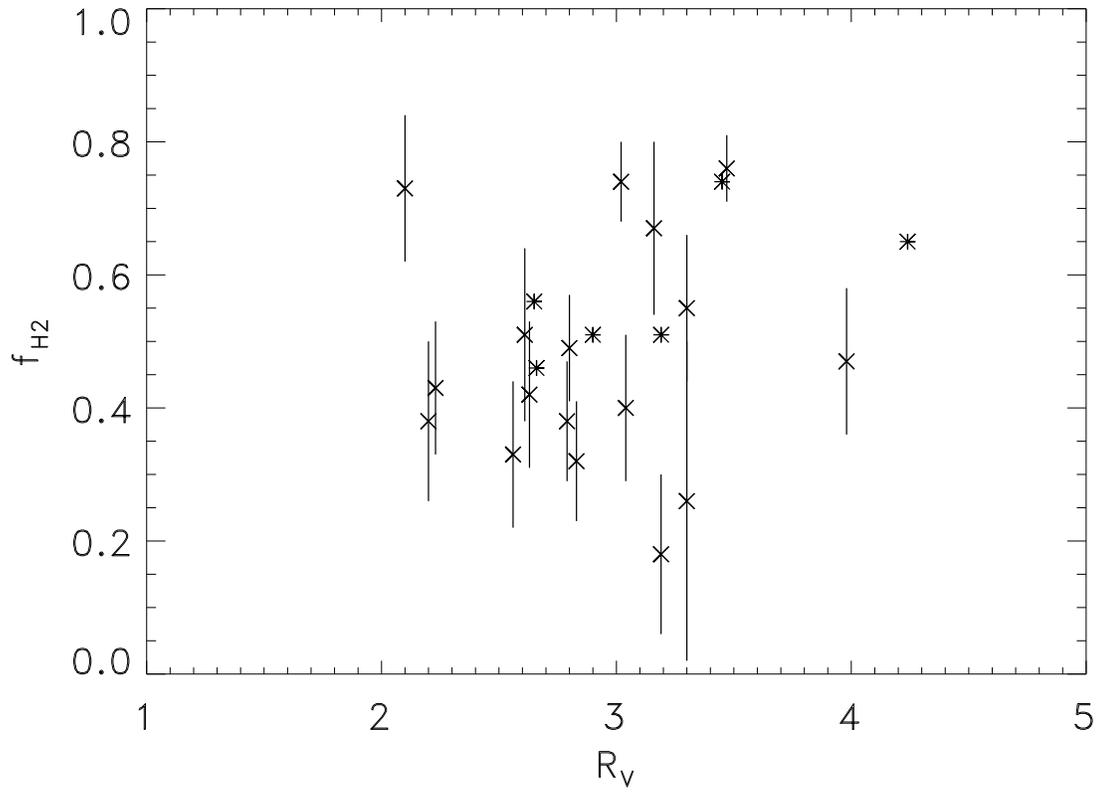}
\caption{Molecular fraction vs. total-to-selective extinction.
Symbols as in Figure 5.}
\end{figure}

\clearpage
\begin{figure}
\plotone{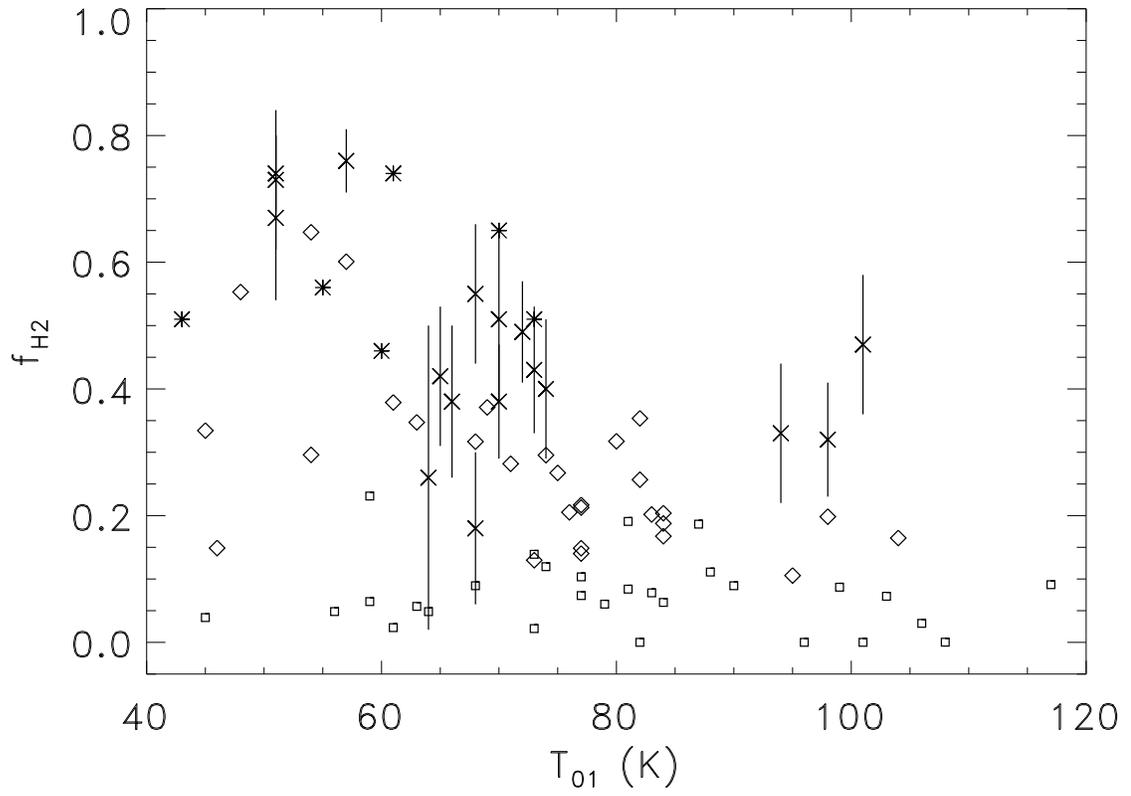}
\caption{Molecular fraction vs. kinetic temperature.  Symbols as in
Figure 7.}
\end{figure}

\begin{figure}
\plotone{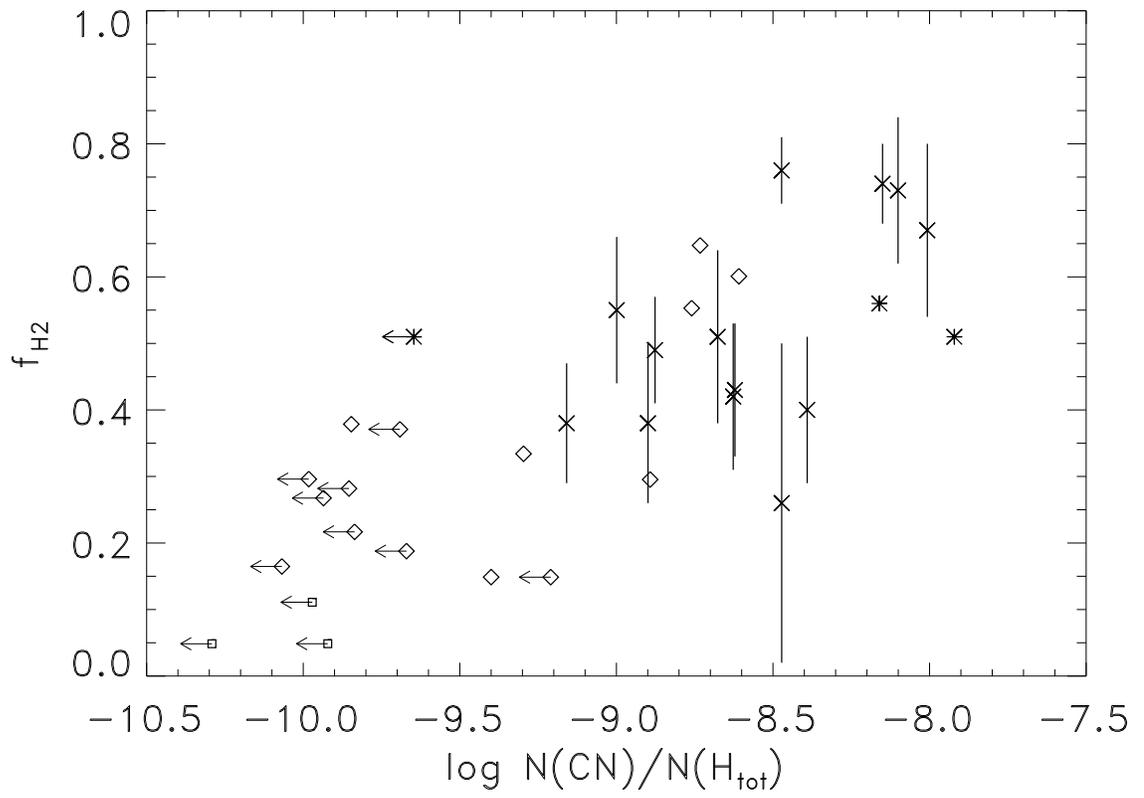}
\caption{Molecular fraction vs. fractional CN abundance.  Symbols as
in Figure 7.}
\end{figure}

\begin{figure}
\plotone{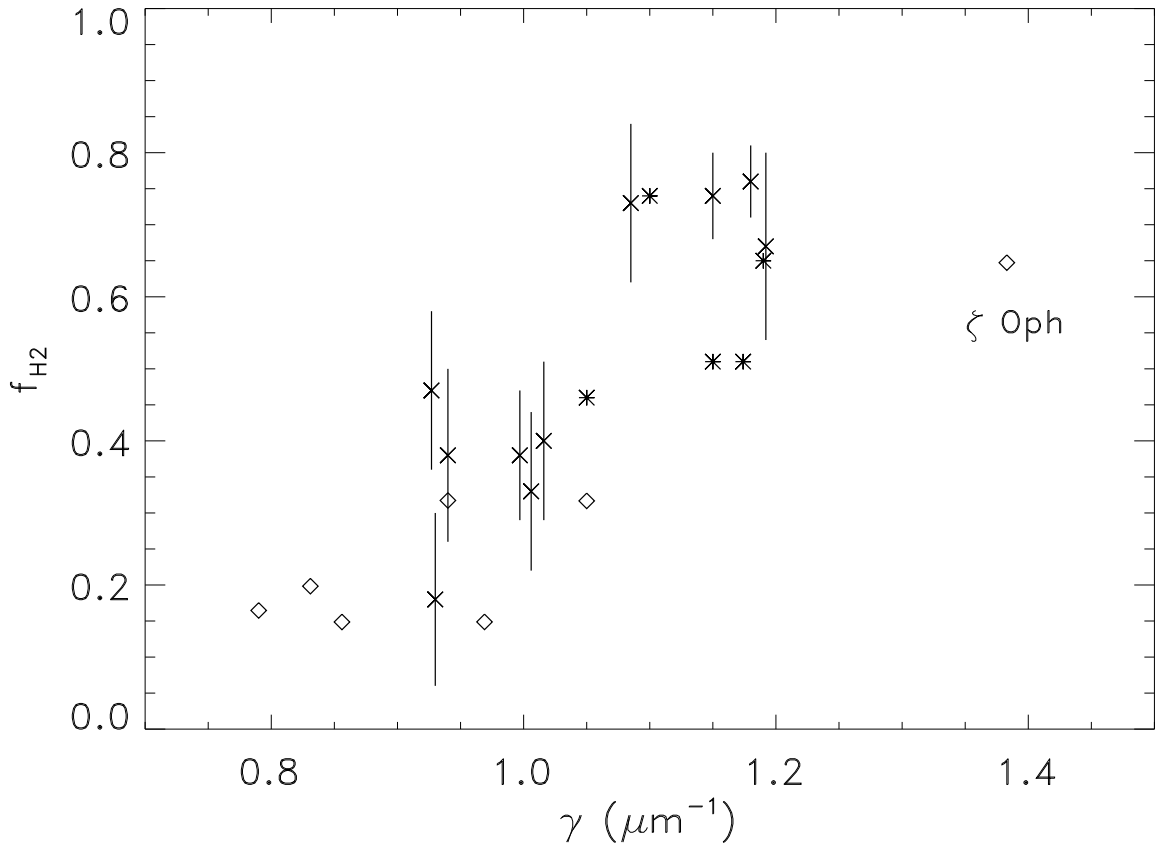}
\caption{Molecular fraction vs. 2175 \AA\ bump width.  Symbols as in
Figure 7; note though that there are no {\it Copernicus} points with
$N$(H$_2$) $<$ 10$^{20}$ cm$^{-3}$.}
\end{figure}

\begin{figure}
\plotone{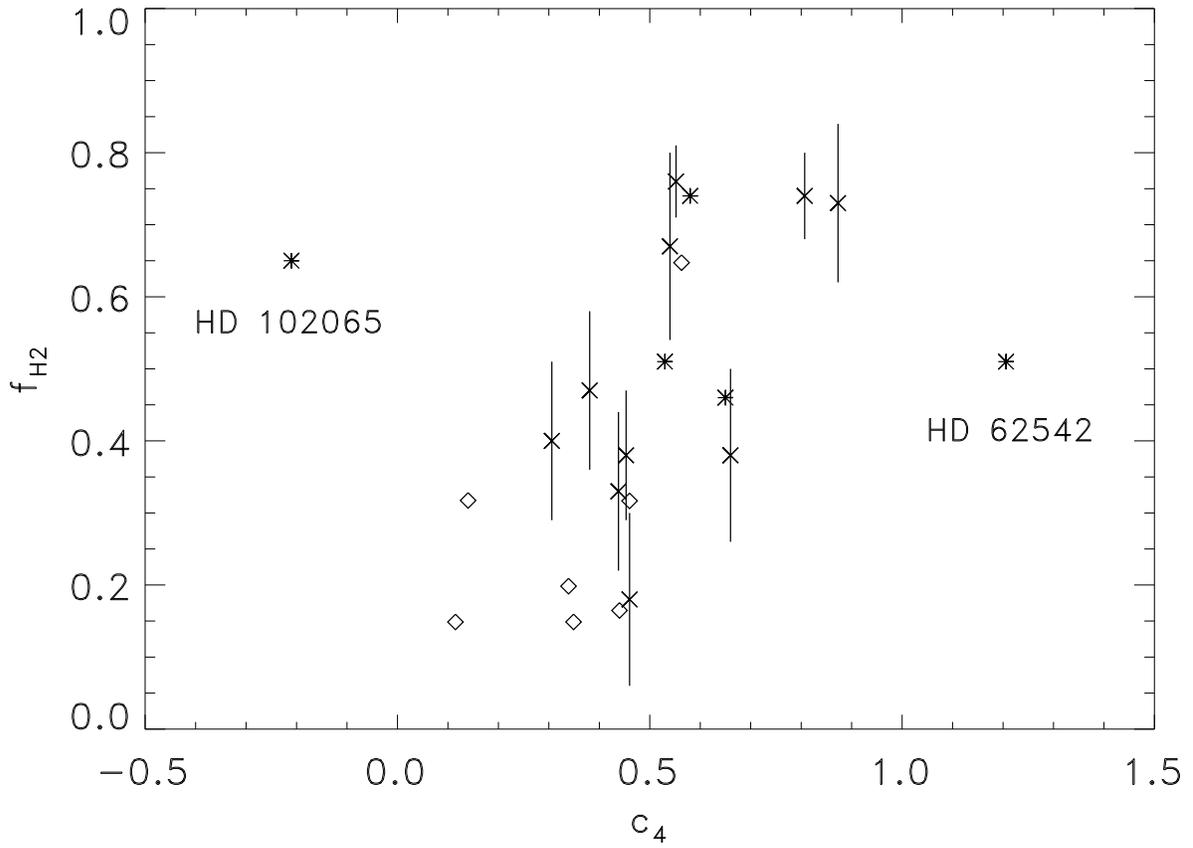}
\caption{Molecular fraction vs. far-UV extinction curvature.  Symbols
as in Figure 7; note though that there are no {\it Copernicus} points with
$N$(H$_2$) $<$ 10$^{20}$ cm$^{-3}$.}
\end{figure}

\clearpage

\begin{deluxetable}{ccccclc}
\tablecaption{Target list}
\tablewidth{0pt}
\tablenum{1}
\tablehead{
  \colhead{Star} & \colhead{$\ell$} & \colhead{$b$} & \colhead{Assoc.}
& \colhead{$V$} & \colhead{MK type} & \colhead{Ref}
}
\startdata
BD $+$31$^{\circ}$ 643 &160.49 &   $-$17.80 &Per OB2 &8.51 &B5 V       &\phn1 \\
HD \phn24534  &163.08    &   $-$17.14 &Per OB2 &6.10 &B0 Ve     &\phn2 \\
HD \phn27778  &172.76    &   $-$17.39 &Tau-Aur &6.49 &B3 V       &\phn3 \\
HD \phn62542  &255.92    &\phn$-$9.24 &        &8.04 &B3 V       &\phn4 \\
HD \phn73882  &260.18    &\phn$+$0.64 &Vel OB1 &7.24 &O8.5 Vn    &\phn5 \\
HD \phn96675  &296.62    &   $-$14.57 &        &7.67 &B7 V       &\phn6 \\
HD 102065 &300.03    &   $-$18.00 &        &6.61 &B9 IV      &\phn7 \\
HD 108927 &301.92    &   $-$15.36 &        &7.78 &B5 V       &\phn7 \\
HD 110432 &301.96    &\phn$-$0.20 &        &5.24 &B1 IIIe    &\phn8 \\
HD 154368 &349.97    &\phn$+$3.22 &        &6.13 &O9.5 Iab   &\phn5 \\
HD 167971 &\phn18.25 &\phn$+$1.68 &Ser OB2 &7.50 &O8 Ib(f)p  &\phn9 \\
HD 168076 &\phn16.94 &\phn$+$0.84 &Ser OB1 &8.21 &O4V((f))   &\phn5 \\
HD 170740 &\phn21.06 &\phn$-$0.53 &        &5.76 &B2 V       &10 \\
HD 185418 &\phn53.60 &\phn$-$2.17 &        &7.45 &B0.5 V     &11 \\
HD 192639 &\phn74.90 &\phn$+$1.48 &Cyg OB1 &7.11 &O7Ib(f)    &12 \\
HD 197512 &\phn87.89 &\phn$+$4.63 &Cyg OB7 &8.57 &B1 V       &12 \\
HD 199579 &\phn85.70 &\phn$-$0.30 &Cyg OB7 &5.96 &O6 V((f))  &12 \\
HD 203938 &\phn90.56 &\phn$-$2.33 &Cyg OB7 &7.09 &B0.5 V     &12 \\
HD 206267 &\phn99.29 &\phn$+$3.74 &Cep OB2 &5.62 &O6.5V((f)) &12 \\
HD 207198 &103.14    &\phn$+$6.99 &Cep OB2 &5.96 &O9.5Ib-II  &12 \\
HD 207538 &101.60    &\phn$+$4.67 &Cep OB2 &7.31 &O9 V       &12 \\
HD 210121 &\phn56.88 &   $-$44.46 &        &7.67 &B3 V       &13 \\
HD 210839 &103.83    &\phn$+$2.61 &Cep OB2 &5.05 &O6 Infp    &12 \\
\enddata
\tablerefs{(1) Harris, Morgan, \& Roman 1954; (2) Roche et al. 1997
  (3) Kenyon et al. 1994; (4) Whittet et al. 1993;
  (5) Walborn 1973; (6) Vrba \& Rydgren 1984;
  (7) Houk 1975; (8) Slettebak 1982;
  (9) Walborn 1972; (10) Murphy 1969;
  (11) Abt 1985; (12) Garmany \& Stencel 1992;
  (13) Welty \& Fowler 1992.
}
\end{deluxetable}

\begin{deluxetable}{ccccccccc}
\tablecaption{Carbon-based molecular abundances}
\tabletypesize{\footnotesize}
\tablewidth{0pt}
\tablenum{2}
\tablehead{
  \colhead{Star} & \colhead{log $N$(CH)} & \colhead{Ref} &
  \colhead{log $N$(CH$^{+}$)} & \colhead{Ref} &
  \colhead{log $N$(CN)} & \colhead{Ref} &
  \colhead{log $N$(CO)} & \colhead{Ref} \\
& \colhead{(cm$^{-2}$)} & & \colhead{(cm$^{-2}$)} & & \colhead{(cm$^{-2}$)}
& & \colhead{(cm$^{-2}$)}
}
\startdata
BD $+$31$^{\circ}$ 643 &13.72 &1
                       &\phm{$<$}13.92 &1
                       &$<$12.04 &1
                       & &\\
HD \phn24534           &13.57 &2
                       &\phm{$<$}13.22 &3
                       &\phm{$<$}12.87 &2
                       &\phm{$<$}16.00 &4 \\
HD \phn27778           &13.48 &5
                       &\phm{$<$}12.85 &5
                       &\phm{$<$}13.18 &5
                       &\phm{$<$}15.82 &6 \\
HD \phn62542           &13.55 &7
                       &$<$11.83 &7
                       &\phm{$<$}13.41 &7
                       &\phm{$<$}16.30\tablenotemark{a} &8 \\
HD \phn73882           &13.57 &9
                       &\phm{$<$}13.38\tablenotemark{b} &9
                       &\phm{$<$}13.58 &10
                       &\phm{$<$}16.48\tablenotemark{a} &8 \\
HD \phn96675           &13.34 &11
                       &\phm{$<$}13.45  &11
                       &       &      
                       &$>$15.00 &11 \\
HD 102065              &12.78 &11
                       &\phm{$<$}13.08  &11
                       &      &
                       &\phm{$<$}13.85 &11 \\
HD 108927 \\
HD 110432              &13.19 &12
                       &\phm{$<$}13.25 &12
                       &\phm{$<$}12.20 &10
                       &\phm{$<$}14.30 &13 \\
HD 154368              &13.80 &2
                       &\phm{$<$}13.67\tablenotemark{b} &3
                       &\phm{$<$}13.26 &2
                       &\phm{$<$}16.00\tablenotemark{a} &8 \\
HD 167971              &13.53 &3
                       &\phm{$<$}13.73 &3 & & & & \\
HD 168076 \\
HD 170740              &13.64 &9
                       &\phm{$<$}13.26 &9
                       &\phm{$<$}12.78 &9
                       &  &\\
HD 185418              &13.12 &2
                       &\phm{$<$}13.10 &2
                       & &
                       & & \\
HD 192639              &13.45 &2
                       &\phm{$<$}13.61 &2
                       & &
                       & & \\
HD 197512 \\
HD 199579              &13.36 &14
                       &\phm{$<$}13.01 &14
                       &\phm{$<$}12.09\tablenotemark{c} &15
                       & & \\
HD 203938              &13.61 &2              
                       &\phm{$<$}13.68\tablenotemark{b} &3
                       &\phm{$<$}13.31 &2
                       & & \\
HD 206267              &13.41\tablenotemark{c} &16
                       &\phm{$<$}13.02 &17
                       &\phm{$<$}12.91 &2
                       & & \\
HD 207198              &13.56 &2
                       &\phm{$<$}13.18 &17
                       &\phm{$<$}12.65 &2
                       &\phm{$<$}15.41\tablenotemark{c} &18 \\
HD 207538              &13.63 &2
                       & &
                       &\phm{$<$}12.96 &2
                       &\phm{$<$}15.15\tablenotemark{a}  &19 \\
HD 210121              &13.48 &20
                       &\phm{$<$}12.78 &21
                       &\phm{$<$}13.09 &10
                       &\phm{$<$}15.48\tablenotemark{b} &20 \\
HD 210839              &13.31 &22
                       &\phm{$<$}13.17 &22
                       &\phm{$<$}12.57 &2
                       &\phm{$<$}15.15\tablenotemark{c} &18 \\
\enddata
\tablenotetext{a}{Emission measurement which should be considered only
an approximation to the column density along the pencil-beam toward the
disk of the star sampled by absorption measurements.}
\tablenotetext{b}{The referenced author(s) assumed $b$ = 2 km s$^{-1}$}
\tablenotetext{c}{The referenced author(s) assumed $b$ = 1 km s$^{-1}$}
\tablerefs{
  (1) Snow et al 1994; (2) D. W. Welty, et al. 2002, in prep.;
  (3) Allen 1994; (4) Kaczmarczyk 2000;
  (5) Meyer \& Roth 1991; (6) Joseph et al. 1986;
  (7) Cardelli et al 1990; (8) van Dishoeck et al 1991;
  (9) Gredel, van Dishoeck, \& Black 1993;
  (10) Gredel, van Dishoeck, \& Black 1991;
  (11) Gry et al. 1998; (12) Crawford 1995;
  (13) Paper II; (14) Jenniskens et al. 1992;
  (15) Joseph, Snow, \& Seab 1989; (16) Federman et al. 1994;
  (17) Chaffee \& Dunham 1979; (18) Federman \& Lambert 1988;
  (19) Dickman et al. 1983; (20) Welty \& Fowler 1992;
  (21) de Vries \& van Dishoeck 1988; (22) Crane, Lambert, \& Sheffer 1995;
}
\end{deluxetable}

\begin{deluxetable}{cccccccccccc}
\tablecaption{Extinction parameters}
\tabletypesize{\footnotesize}
\tablewidth{0pt}
\tablenum{3}
\tablehead{
& & & \multicolumn{8}{c}{$R_V$} \\
\cline{4-11} \\
  \colhead{Star} & \colhead{$E(B-V)$} & \colhead{Ref}
& \colhead{Phot.\tablenotemark{a}} &\colhead{Filters} & \colhead{Ref}
& \colhead{Polar.\tablenotemark{b}} & \colhead{Ref}
& \colhead{E.C.\tablenotemark{c}} & \colhead{Ref}
& \colhead{Adopted} & \colhead{$A_V$}
}
\startdata
BD $+$31$^{\circ}$ 643 &0.84 &\phn1 &3.19 &$HKL$  &\phn2 &3.75 &\phn3 &3.46 &\phn1 &3.19 &2.68 \\
HD \phn24534  &0.45 &\phn4 &     &      &  &3.47 &\phn5 &3.20 &\phn4 &3.47 &1.56 \\
HD \phn27778  &0.38 &\phn6 &2.65 &$JK$   &\phn6 &     &  &     &  &2.65 &1.01 \\
HD \phn62542  &0.37 &\phn7 &2.90 &$JHKL$ &\phn7 &3.27 &\phn8 &2.14 &\phn9 &2.90 &1.07 \\
HD \phn73882  &0.72 &10 &3.16 &$JHKL$ &11 &3.51 &\phn8 &2.93 &10 &3.16 &2.28 \\
HD \phn96675  &0.31 &12 &3.45 &$JHKL$ &12 &2.80 &13    &4.02 &12 &3.45 &1.07 \\
HD 102065 &0.17 &12 &4.24 &$JHKL$ &12 &     &      &3.36 &12 &4.24 &0.72 \\
HD 108927 &0.23 &12 &2.66 &$JHK$  &12 &     &      &3.73 &12 &2.66 &0.61 \\
HD 110432 &0.40 &14 &\tablenotemark{d} &$JHKL$ &11 &3.30 &15 & & &3.30 &1.32 \\
HD 154368 &0.82 &16 &3.02 &$JHK$  &17 &     &      &3.14 &16 &3.02 &2.48 \\
HD 167971 &1.04 &18 &3.30 &$JHKL$ &18 &     &      &     &   &3.30 &3.43 \\
HD 168076 &0.79 &19 &\tablenotemark{d} &$JHKL$ &19 &3.19 &20 &3.62 &21 &3.19 &2.86 \\
HD 170740 &0.48 &22 &2.61 &$JHKL$ &11 &3.08 &15    &     &   &2.61 &1.25 \\
HD 185418 &0.51 &23 &     &       &   &     &      &3.98 &10 &3.98 &2.03 \\
HD 192639 &0.66 &24 &2.83 &$JHKLM$ &25 &    &      &     &   &2.83 &1.87 \\
HD 197512 &0.33 &24 &     &       &   &     &      &2.56 &10 &2.56 &0.84 \\
HD 199579 &0.36 &24 &2.79 &$KL$   &26 &     &      &2.74 &10 &2.79 &1.00 \\
HD 203938 &0.72 &24 &3.04 &$KLM$  &26 &     &      &3.00 &10 &3.04 &2.19 \\
HD 206267 &0.52 &24 &2.63 &$JHK$  &27 &     &      &     &   &2.63 &1.37 \\
HD 207198 &0.62 &24 &2.20 &$JK$   &19 &2.30 &28    &2.66 &21 &2.20 &1.36 \\
HD 207538 &0.64 &24 &     &       &   &2.23 &\phn8 &     &   &2.23 &1.43 \\
HD 210121 &0.38 &29 &2.10 &$JHKL$ &29 &2.13 &29 &2.01 &\phn9 &2.10 &0.80 \\
HD 210839 &0.56 &24 &2.80 &$JHKLM$ &25 &2.86 &30 & &           &2.80 &1.57 \\
\enddata
\tablenotetext{a}{Derived from Eq. 1 using infrared photometry from the given reference}
\tablenotetext{b}{Derived from the wavelength of maximum polarization;
$R_V$ = 5.6$\lambda_{\rm max}$ with $\lambda_{\rm max}$ in $\mu$m}.
\tablenotetext{c}{Derived from the linear far-UV rise in the extinction curve}
\tablenotetext{d}{No reasonable solution could be obtained}
\tablerefs{(1) Snow et al. 1994; (2) Strom, Strom, \& Carrasco 1974;
  (3) Andersson \& Wannier 2000; (4) Snow et al. 1998;
  (5) Roche et al. 1997; (6) Kenyon, Dobrzycka, \& Hartmann 1994;
  (7) Whittet et al. 1993; (8) Martin, Clayton, \& Wolff 1999;
  (9) Welty \& Fowler 1992; (10) Fitzpatrick \& Massa 1990;
  (11) Whittet \& van Breda 1978; (12) Boulanger et al. 1994;
  (13) Whittet et al. 1994; (14) Paper II;
  (15) Serkowski, Mathewson, Ford 1975; (16) Snow et al. 1996;
  (17) Th\'{e}, Wesselius, \& Jannsen 1986; (18) Leitherer et al. 1987;
  (19) Aiello et al. 1988; (20) Orsatti, Vega, \& Marraco 2000;
  (21) Jenniskens \& Greenberg 1993; (22) Clayton \& Mathis 1988;
  (23) S\={u}d\v{z}ias \& Bobinas 1984; (24) Garmany \& Stencel 1992;
  (25) Castor \& Simon 1983; (26) Sneden et al. 1978;
  (27) Morbidelli et al. 1997; (28) Anderson et al. 1996;
  (29) Larson, Whittet, \& Hough 1996; (30) McDavid 2000
}
\end{deluxetable}

\begin{deluxetable}{cccccccc}
\tablecaption{Extinction curve parameters\tablenotemark{a}}
\tablewidth{0pt}
\tablenum{4}
\tablehead{
  \colhead{Target} & \colhead{$\lambda_0^{-1}$} & \colhead{$\gamma$}
& \colhead{c$_1$} & \colhead{c$_2$} & \colhead{c$_3$} & \colhead{c$_4$}
& \colhead{Ref.} \\
& \colhead{($\mu$m$^{-1}$)} & \colhead{($\mu$m$^{-1}$)}
}
\startdata
BD $+$31$^{\circ}$ 643 &4.613    &1.28\phn &\phm{$-$}0.39\phn &0.51\phn &5.37\phn  &\phm{$-$}0.58\phn &1 \\
HD \phn24534           &         &1.18\phn &                  &         &          &\phm{$-$}0.552    &2 \\
HD \phn27778  \\
HD \phn62542           &4.723    &1.174    &$-$2.416          &1.383    &1.949     &\phm{$-$}1.206    &3 \\
HD \phn73882           &4.576    &1.192    &$-$0.412          &0.788    &3.341     &\phm{$-$}0.540    &4 \\
HD \phn96675           &4.63\phn &1.10\phn &\phm{$-$}0.99\phn &0.35\phn &4.31\phn  &\phm{$-$}0.58\phn &5 \\
HD 102065              &4.59\phn &1.19\phn &$-$1.50\phn       &0.81\phn &8.00\phn  &$-$0.21\phn &5 \\
HD 108927              &4.67\phn &1.05\phn &\phm{$-$}1.48\phn &0.44\phn &4.00\phn  &\phm{$-$}0.65\phn &5 \\
HD 110432  \\
HD 154368              &4.581    &1.15\phn &$-$0.01\phn       &0.680    &4.419     &\phm{$-$}0.807    &6 \\
HD 167971  \\
HD 168076              &4.595    &0.93\phn &\phm{$-$}0.57\phn &0.48\phn &2.85\phn  &\phm{$-$}0.46\phn &1 \\
HD 170740  \\
HD 185418              &4.579    &0.927    &\phm{$-$}1.266    &0.362    &3.941     &\phm{$-$}0.381    &4 \\
HD 192639  \\
HD 197512              &4.585    &1.006    &$-$1.043          &1.021    &4.659     &\phm{$-$}0.438    &4 \\
HD 199579              &4.606    &0.997    &$-$0.725          &0.898    &2.923     &\phm{$-$}0.453    &4 \\
HD 203938              &4.589    &1.016    &\phm{$-$}0.087    &0.747    &3.647     &\phm{$-$}0.306    &4 \\
HD 206267  \\
HD 207198              &4.596    &0.94\phn &$-$0.91\phn       &0.95\phn &3.13\phn  &\phm{$-$}0.66\phn &1 \\
HD 207538 \\
HD 210121              &4.603    &1.085    &$-$2.493          &1.528    &2.405     &\phm{$-$}0.873    &3 \\
HD 210839 \\
\enddata
\tablenotetext{a}{In the parameterization scheme of Fitzpatrick \& Massa 1990}
\tablerefs{(1) Jenniskens \& Greenberg 1993; (2) Snow et al. 1998;
   (3) Welty \& Fowler 1992; (4) Fitzpatrick \& Massa 1990;
   (5) Boulanger et al. 1994; (6) Snow et al. 1996}
\end{deluxetable}

\begin{deluxetable}{clccc}
\tablecaption{{\it FUSE} observations}
\tablewidth{0pt}
\tablenum{5}
\tablehead{
  \colhead{Target} & \colhead{Date} &
  \colhead{N$_{\rm int}$\tablenotemark{a}} &
  \colhead{t$_{\rm int}$\tablenotemark{b}} &
  \colhead{S/N\tablenotemark{c}} \\
& & & \colhead{(ksec)} & 
}
\startdata
BD $+$31$^{\circ}$ 643 &1999 Nov 13 &    17 &    37.5 & \phn1.0 \\
\ldots       &2000 Oct 26 &    13 &    34.7 & \phn1.1 \\
HD \phn24534     &2000 Sep 12 & \phn8 & \phn8.3 &    24.3 \\
HD \phn27778     &2000 Oct 27 & \phn4 & \phn9.7 &    10.8 \\
HD \phn62542     &2000 Jan 25 & \phn5 &    11.0 & \phn3.1 \\
HD \phn73882     &1999 Oct 30 &    11 &    25.5 & \phn5.1 \\
\ldots       &2000 Jan 24 & \phn6 &    11.9 & \phn4.7 \\
\ldots       &2000 Mar 19 & \phn8 &    13.6 & \phn4.6 \\
HD \phn96675     &2000 Apr 10 & \phn9 &    10.2 & \phn8.4 \\
HD 102065    &2000 May 28 & \phn6 & \phn6.8 & \phn7.3 \\
HD 108927    &2000 May 27 & \phn2 & \phn6.5 &    11.2 \\
HD 110432    &2000 Apr 4  & \phn5 & \phn3.6 &    28.5 \\
HD 154368    &2000 Jul 14 & \phn8 &    12.5 & \phn2.3 \\
HD 167971    &2000 May 16 & \phn3 & \phn9.5 & \phn2.0 \\
HD 168076    &2000 May 16 & \phn2 & \phn6.6 & \phn5.9 \\
HD 170740    &2001 Apr 27 & \phn5 & \phn2.9 & \phn8.8 \\
HD 185418    &2000 Aug 10 & \phn3 & \phn4.4 &    14.9 \\
HD 192639    &2000 Jun 12 & \phn2 & \phn4.8 & \phn8.1 \\
HD 197512    &2001 Jun 6  & \phn3 & \phn8.2 &    10.9 \\
HD 199579    &2000 Jul 19 & \phn8 & \phn4.3 &    29.2 \\
HD 203938    &2000 Jul 20 & \phn4 & \phn7.8 & \phn3.5 \\
HD 206267    &2000 Jul 21 & \phn3 & \phn4.9 &    10.2 \\
HD 207198    &2000 Jul 23 & \phn3 &    13.2 &    11.2 \\
HD 207538    &1999 Dec 8  & \phn7 & \phn7.7 & \phn6.2 \\
\ldots       &2000 Jul 21 &    10 &    11.2 & \phn7.1 \\
HD 210121    &2000 Jun 29 & \phn5 &    13.8 & \phn2.9 \\
HD 210839    &2000 Jul 22 &    10 & \phn6.1 &    24.0 \\
\enddata
\tablenotetext{a}{Number of integrations}
\tablenotetext{b}{Total integration time}
\tablenotetext{c}{Average per-pixel S/N for a 1 \AA\ region of the LiF 1A
spectrum near 1070 \AA , between the Lyman (3,0) and (2,0) bandheads of H$_2$.
One resolution element corresponds to about 9 pixels.}
\end{deluxetable}

\begin{deluxetable}{ccccccc}
\tablecaption{Fitting test for HD 199579}
\tablewidth{0pt}
\tablenum{6}
\tablehead{
  \colhead{Segment} & \colhead{Band} & \colhead{Method} & \colhead{Weights}
& \colhead{log $N$(0)} & \colhead{log $N$(1)} \\
& & & & \colhead{(cm $^{-2}$)} & \colhead{(cm $^{-2}$)}}
\startdata
LiF 1A &(2,0) & CURFIT & None      &20.291 &20.155 \\
LiF 1A &(2,0) & AMOEBA & None      &20.290 &20.154 \\
LiF 1A &(2,0) & CURFIT & S/N       &20.286 &20.154 \\
LiF 1A &(2,0) & AMOEBA & S/N       &20.285 &20.153 \\
LiF 1A &(2,0) & CURFIT & $\chi^2$  &20.308 &20.157 \\
LiF 1A &(2,0) & AMOEBA & $\chi^2$  &20.306 &20.155 \\

LiF 1A &(4,0) & CURFIT & None      &20.319 &20.106 \\
LiF 1A &(4,0) & AMOEBA & None      &20.319 &20.105 \\
LiF 1A &(4,0) & CURFIT & S/N       &20.314 &20.092 \\
LiF 1A &(4,0) & AMOEBA & S/N       &20.314 &20.091 \\
LiF 1A &(4,0) & CURFIT & $\chi^2$  &20.336 &20.134 \\
LiF 1A &(4,0) & AMOEBA & $\chi^2$  &20.335 &20.132 \\

LiF 2A &(1,0) & CURFIT & None      &20.314 &20.138 \\
LiF 2A &(1,0) & AMOEBA & None      &20.316 &20.141 \\
LiF 2A &(1,0) & CURFIT & S/N       &20.323 &20.144 \\
LiF 2A &(1,0) & AMOEBA & S/N       &20.321 &20.141 \\
LiF 2A &(1,0) & CURFIT & $\chi^2$  &20.297 &20.123 \\
LiF 2A &(1,0) & AMOEBA & $\chi^2$  &20.289 &20.116 \\

\enddata
\end{deluxetable}

\begin{deluxetable}{cccc}
\tablecaption{Band-to-band variation}
\tablewidth{0pt}
\tablenum{7}
\tablehead{
  \colhead{Segment} & \colhead{Band}
& \colhead{$d_{0}$\tablenotemark{a}}
& \colhead{$d_{1}$\tablenotemark{a}}
}
\startdata
LiF 1A  &(2,0) &$-$1.44$\pm$0.15 &$-$0.39$\pm$0.21 \\
LiF 1A  &(4,0) &$+$0.52$\pm$0.17 &$-$0.40$\pm$0.23 \\
LiF 2A  &(1,0) &$+$0.93$\pm$0.16 &$-$0.36$\pm$0.25 \\
LiF 2B  &(4,0) &$-$0.29$\pm$0.16 &$+$0.40$\pm$0.18 \\
SiC 1A  &(2,0) &$-$0.12$\pm$0.17 &$+$0.24$\pm$0.26 \\
SiC 1A  &(4,0) &$-$0.30$\pm$0.17 &$+$0.38$\pm$0.17 \\
SiC 2B  &(1,0) &$+$0.68$\pm$0.22 &$-$0.40$\pm$0.20 \\
SiC 2B  &(2,0) &$-$0.14$\pm$0.20 &$+$0.55$\pm$0.25 \\
SiC 2B  &(4,0) &$+$0.17$\pm$0.17 &$-$0.03$\pm$0.24 \\
All     &(1,0) &$+$0.80$\pm$0.13 &$-$0.38$\pm$0.16 \\
All     &(2,0) &$-$0.56$\pm$0.13 &$+$0.15$\pm$0.14 \\
All     &(4,0) &$+$0.02$\pm$0.09 &$+$0.09$\pm$0.11 \\
All     &All &\phm{$-$}0.00$\pm$0.08 &\phm{$-$}0.00$\pm$0.08 \\
\enddata
\tablenotetext{a}{Average normalized deviation from the mean and
error of the mean
($\sigma_{n-1}$/$\sqrt{n}$), for $J$=0 and $J$=1 measurements}
\end{deluxetable}

\begin{deluxetable}{cccccccccc}
\tablecaption{Molecular and atomic hydrogen parameters}
\tabletypesize{\footnotesize}
\tablewidth{0pt}
\tablenum{8}
\tablehead{
  \colhead{Target} & \colhead{Bands} & \colhead{log $N$(H$_2$)}
& \colhead{log $N$(0)}
& \colhead{log $N$(1)} & \colhead{T$_{kin}$} & \colhead{log $N$(H I)}
& \colhead{Ref} & \colhead{f$_{H2}$} \\
& & \colhead{(cm$^{-2}$)} & \colhead{(cm$^{-2}$)} & \colhead{(cm$^{-2}$)}
& \colhead{(K)} &  \colhead{(cm$^{-2}$)} & &
}
\startdata
BD $+$31$^{\circ}$ 643 &\phn3 &21.09$\pm$0.19 &20.82$\pm$0.16 &20.76$\pm$0.24
          &\phn73$\pm$48     &21.38$\pm$0.30  &1 &0.51$\pm$0.26 \\
HD \phn24534  &\phn9  &20.92$\pm$0.04  &20.76$\pm$0.03  &20.42$\pm$0.06
          &\phn57$\pm$\phn4  &20.73$\pm$0.06  &2 &0.76$\pm$0.05 \\
HD \phn27778  &\phn9  &20.79$\pm$0.06  &20.64$\pm$0.05  &20.27$\pm$0.10
          &\phn55$\pm$\phn7  &20.98$\pm$0.30  &1 &0.56$\pm$0.20 \\
HD \phn62542  &\phn3  &20.81$\pm$0.21  &20.74$\pm$0.21  &19.98$\pm$0.14
          &\phn43$\pm$11     &20.93$\pm$0.30  &1 &0.60$\pm$0.28 \\
HD \phn73882  &10     &21.11$\pm$0.08  &20.99$\pm$0.08  &20.50$\pm$0.07
          &\phn51$\pm$\phn6  &21.11$\pm$0.15  &3 &0.67$\pm$0.13 \\
HD \phn96675  &\phn9  &20.82$\pm$0.05  &20.63$\pm$0.04  &20.37$\pm$0.08
          &\phn61$\pm$\phn7  &20.66$\pm$0.30  &1 &0.74$\pm$0.18 \\
HD 102065 &\phn9  &20.50$\pm$0.06  &20.25$\pm$0.06  &20.15$\pm$0.06
          &\phn70$\pm$\phn9  &20.54$\pm$0.30  &1 &0.65$\pm$0.21 \\
HD 108927 &\phn9  &20.49$\pm$0.09  &20.30$\pm$0.09  &20.03$\pm$0.09
          &\phn60$\pm$10     &20.86$\pm$0.30  &1 &0.46$\pm$0.21 \\
HD 110432 &\phn9  &20.64$\pm$0.04  &20.40$\pm$0.03  &20.27$\pm$0.04
          &\phn68$\pm$\phn5  &20.85$\pm$0.15  &4 &0.55$\pm$0.11 \\
HD 154368 &\phn7  &21.16$\pm$0.07  &21.04$\pm$0.05  &20.54$\pm$0.15
          &\phn51$\pm$\phn8  &21.00$\pm$0.05  &5 &0.74$\pm$0.06 \\
HD 167971 &\phn4  &20.85$\pm$0.12  &20.64$\pm$0.10  &20.44$\pm$0.15
          &\phn64$\pm$17     &21.60$\pm$0.30  &6 &0.26$\pm$0.22 \\
HD 168076 &\phn9  &20.68$\pm$0.08  &20.44$\pm$0.08  &20.31$\pm$0.09
          &\phn68$\pm$13     &21.65$\pm$0.23  &2 &0.18$\pm$0.12 \\
HD 170740 &\phn7  &20.86$\pm$0.08  &20.60$\pm$0.05  &20.52$\pm$0.11
          &\phn70$\pm$13     &21.15$\pm$0.15  &2 &0.51$\pm$0.13 \\
HD 185418 &\phn9  &20.76$\pm$0.05  &20.34$\pm$0.04  &20.56$\pm$0.05
          &101$\pm$14    &21.11$\pm$0.15  &3 &0.47$\pm$0.11 \\
HD 192639 &\phn9  &20.69$\pm$0.05  &20.28$\pm$0.05  &20.48$\pm$0.05
          &\phn98$\pm$15     &21.32$\pm$0.12  &2 &0.32$\pm$0.09 \\
HD 197512 &\phn9  &20.66$\pm$0.05  &20.27$\pm$0.05  &20.44$\pm$0.05
          &\phn94$\pm$14     &21.26$\pm$0.15  &3 &0.33$\pm$0.11 \\
HD 199579 &\phn9  &20.53$\pm$0.04  &20.28$\pm$0.03  &20.17$\pm$0.03
          &\phn70$\pm$\phn5  &21.04$\pm$0.11  &2 &0.38$\pm$0.09 \\
HD 203938 &\phn6  &21.00$\pm$0.06  &20.72$\pm$0.05  &20.68$\pm$0.08
          &\phn74$\pm$\phn9  &21.48$\pm$0.15  &3 &0.40$\pm$0.11 \\
HD 206267 &\phn9  &20.86$\pm$0.04  &20.64$\pm$0.03  &20.45$\pm$0.05
          &\phn65$\pm$\phn5  &21.30$\pm$0.15  &6 &0.42$\pm$0.11 \\
HD 207198 &\phn9  &20.83$\pm$0.04  &20.61$\pm$0.03  &20.44$\pm$0.04
          &\phn66$\pm$\phn5  &21.34$\pm$0.17  &2 &0.38$\pm$0.12 \\
HD 207538 &18     &20.91$\pm$0.06  &20.64$\pm$0.07  &20.58$\pm$0.05
          &\phn73$\pm$\phn8  &21.34$\pm$0.12  &2 &0.43$\pm$0.10 \\
HD 210121 &\phn5  &20.75$\pm$0.12  &20.63$\pm$0.11  &20.13$\pm$0.15
          &\phn51$\pm$11     &20.63$\pm$0.15  &7 &0.73$\pm$0.11 \\
HD 210839 &\phn9  &20.84$\pm$0.04  &20.57$\pm$0.04  &20.50$\pm$0.04
          &\phn72$\pm$\phn6  &21.15$\pm$0.10  &2 &0.49$\pm$0.08 \\
\enddata
\tablerefs{(1) Present work; $N$(H I) = 5.8$\times$10$^{21}$$E(B-V)$ -- 2$N$(H$_2)$;
   (2) Diplas \& Savage 1994; (3) Fitzpatrick \& Massa 1990;
   (4) Paper II; (5) Snow et al. 1996;
   (6) Present work; Ly$\alpha$ profile fitting;
   (7) Welty \& Fowler 1992; 21-cm emission measurement with possible
systematic errors relative to the absorption measures}
\end{deluxetable}

\begin{deluxetable}{cccccccccc}
\tablecaption{Comparison with {\it Copernicus} results}
\tablewidth{0pt}
\tablenum{9}
\tablehead{
  \colhead{Target} & \colhead{Satellite} & \colhead{log $N$(H$_2$)}
& \colhead{log $N$(0)} & \colhead{log $N$(1)} & \colhead{Ref.} \\
& & \colhead{(cm$^{-2}$)} & \colhead{(cm$^{-2}$)} & \colhead{(cm$^{-2}$)}
}
\startdata
HD \phn24534  & {\it FUSE}   & 20.92$\pm$0.04 & 20.76$\pm$0.03 & 20.42$\pm$0.06 & \\
\ldots & {\it Copernicus} & 21.04$\pm$0.13 & 20.78$\pm$0.11 & 20.70$\pm$0.13 & 1 \\
HD 199579 & {\it FUSE}   & 20.53$\pm$0.04 & 20.28$\pm$0.03 & 20.17$\pm$0.17 & \\
\ldots & {\it Copernicus} & 20.36$\pm$0.18 & 20.08$\pm$0.18 & 20.04$\pm$0.18 & 2 \\
HD 210839 & {\it FUSE}   & 20.84$\pm$0.04 & 20.57$\pm$0.04 & 20.50$\pm$0.04 & \\
\ldots & {\it Copernicus} & 20.78          &                &   & 3 \\
\enddata
\tablerefs{
  (1) Mason et al. 1976; (2) Savage et al. 1977; (3) Bohlin et al. 1978 }
\end{deluxetable}

\begin{deluxetable}{ccccc}
\tablecaption{CH component structure\tablenotemark{a}}
\tablewidth{0pt}
\tablenum{10}
\tablehead{
  \colhead{Target} & \colhead{R\tablenotemark{b}} &
  \colhead{N$_{comp}$\tablenotemark{c}} &
  \colhead{Max$_{comp}$\tablenotemark{d}} & \colhead{$f_{\rm H2}$} \\
& \colhead{(km s$^{-1}$)}
}
\startdata
BD $+$31$^{\circ}$ 643 &3.6 &1 &1.00 &0.51 \\
HD \phn24534           &1.7 &2 &0.72 &0.76 \\
HD \phn27778           &1.3 &2 &0.53 &0.56 \\
HD \phn73882           &3.6 &1 &1.00 &0.67 \\
HD 110432              &0.3 &2 &0.68 &0.55 \\
HD 154368              &1.5 &4 &0.78 &0.74 \\
HD 167971              &1.5 &4 &0.47 &0.26 \\
HD 170740              &2.5 &1 &1.00 &0.51 \\
HD 185418              &1.7 &2 &0.59 &0.47 \\
HD 192639              &1.7 &2 &0.56 &0.32 \\
HD 199579              &1.7 &5 &0.64 &0.38 \\
HD 203938              &1.7 &3 &0.45 &0.40 \\
HD 206267              &1.3 &3 &0.50 &0.42 \\
HD 207198              &1.3 &3 &0.74 &0.38 \\
HD 207538              &2.0 &4 &0.39 &0.43 \\
HD 210121              &3.6 &1 &1.00 &0.73 \\
HD 210839              &0.6 &2 &0.62 &0.49 \\
\enddata
\tablenotetext{a}{Data from D. E. Welty et al 2002, in preparation,
except Crawford 1995 for HD 110432 and Crane et al. 1995 for
HD 210839}
\tablenotetext{b}{Spectral resolution}
\tablenotetext{c}{Number of components}
\tablenotetext{d}{Fractional abundance of strongest component}
\end{deluxetable}

\begin{deluxetable}{ccccccccc}
\tablecaption{Separation into ``translucent'' and ``diffuse''
components\tablenotemark{a}}
\tablewidth{0pt}
\tabletypesize{\scriptsize}
\tablenum{11}
\tablehead{
  \colhead{Target} & \colhead{$E(B-V)$}
& \colhead{H$_{\rm tot}$} & \colhead{$f_{\rm H2}$} & \colhead{$T_{01}$}
& \colhead{log $N$(H I)} & \colhead{log $N$(H$_2$)} & \colhead{log $N$(0)}
& \colhead{log $N$(1)} \\
& & \colhead{(cm$^{-2}$)} & & \colhead{(K)} & \colhead{(cm$^{-2}$)}
& \colhead{(cm$^{-2}$)} & \colhead{(cm$^{-2}$)} & \colhead{(cm$^{-2}$)}
}
\startdata
BD+31 643 &0.20/0.64 &21.06/21.57 &0.90/0.38 &30/187 &20.06/21.36 &20.72/20.85 &20.70/20.19 &19.19/20.75 \\

HD  \phn24534 &0.20/0.25 &21.06/21.02 &0.90/0.60 &30/185 &20.06/20.62 &20.72/20.50 &20.70/19.84 &19.19/20.39 \\

HD  \phn27778 &0.15/0.23 &20.94/21.12 &0.90/0.35 &30/158 &19.94/20.94 &20.59/20.36 &20.58/19.75 &19.07/20.24 \\

HD  \phn62542 &0.20/0.17 &21.06/20.99 &0.90/0.25 &30/107 &20.06/20.87 &20.72/20.09 &20.70/19.63 &19.19/19.90 \\

HD  \phn73882 &0.35/0.37 &21.31/21.26 &0.90/0.41 &30/163 &20.31/21.04 &20.96/20.58 &20.95/19.96 &19.43/20.46 \\

HD  \phn96675 &0.14/0.17 &20.91/20.99 &0.90/0.61 &30/160 &19.91/20.58 &20.56/20.47 &20.55/19.86 &19.04/20.35 \\

HD 102065 &0.05/0.12 &20.46/20.84 &0.90/0.54 &30/140 &19.46/20.50 &20.12/20.28 &20.10/19.74 &18.59/20.14 \\

HD 108927 &0.06/0.17 &20.54/21.00 &0.90/0.30 &30/119 &19.54/20.84 &20.19/20.18 &20.18/19.68 &18.67/20.01 \\

HD 110432 &0.08/0.32 &20.67/21.05 &0.90/0.41 &30/191 &19.67/20.82 &20.32/20.36 &20.31/19.69 &18.79/20.26 \\

HD 154368 &0.40/0.42 &21.37/21.20 &0.90/0.51 &30/196 &20.37/20.89 &21.02/20.60 &21.01/19.92 &19.49/20.50 \\

HD 167971 &0.14/0.90 &20.91/21.66 &0.90/0.15 &30/166 &19.91/21.59 &20.56/20.54 &20.55/19.91 &19.04/20.42 \\

HD 168076 &0.08/0.71 &20.67/21.70 &0.90/0.11 &30/142 &19.67/21.65 &20.32/20.43 &20.31/19.86 &18.79/20.30 \\

HD 170740 &0.12/0.36 &20.84/21.34 &0.90/0.38 &30/175 &19.84/21.13 &20.50/20.62 &20.48/19.97 &18.97/20.51 \\

HD 185418 &0.04/0.47 &20.37/21.34 &0.90/0.43 &30/158 &19.37/21.10 &20.02/20.68 &20.01/20.07 &18.49/20.56 \\

HD 192639 &0.04/0.62 &20.37/21.45 &0.90/0.27 &30/172 &19.37/21.32 &20.02/20.59 &20.01/19.95 &18.49/20.48 \\

HD 197512 &0.04/0.29 &20.37/21.40 &0.90/0.28 &30/165 &19.37/21.25 &20.02/20.55 &20.01/19.93 &18.49/20.44 \\

HD 199579 &0.06/0.30 &20.54/21.15 &0.90/0.26 &30/192 &19.54/21.03 &20.19/20.26 &20.18/19.59 &18.67/20.16 \\

HD 203938 &0.15/0.57 &20.94/21.62 &0.90/0.29 &30/165 &19.94/21.47 &20.59/20.79 &20.58/20.16 &19.07/20.67 \\

HD 206267 &0.14/0.38 &20.91/21.42 &0.90/0.27 &30/170 &19.91/21.28 &20.56/20.55 &20.55/19.91 &19.04/20.43 \\

HD 207198 &0.13/0.49 &20.88/21.44 &0.90/0.25 &30/174 &19.88/21.32 &20.53/20.45 &20.52/19.89 &19.00/20.42 \\

HD 207538 &0.13/0.51 &20.88/21.49 &0.90/0.31 &30/177 &19.88/21.32 &20.53/20.68 &20.52/20.03 &19.00/20.57 \\

HD 210121 &0.15/0.23 &20.94/20.83 &0.90/0.50 &30/138 &19.94/20.53 &20.59/20.23 &20.58/19.67 &19.07/20.09 \\

HD 210839 &0.11/0.45 &20.80/21.33 &0.90/0.37 &30/170 &19.80/21.13 &20.46/20.60 &20.44/19.97 &18.93/20.49 \\

\enddata
\tablenotetext{a}{In each pair of values, the value for the ``translucent'' component is given first.}
\end{deluxetable}

\end{document}